\documentclass[11pt,a4paper]{article}
\pdfoutput=1
\usepackage{jheppub}
\usepackage{graphicx,psfrag,color}% Include figure files
\usepackage{bm}% bold math
\usepackage{mathbbol,verbatim, mathrsfs}
\usepackage{slashed}
\usepackage{braket}
\usepackage{graphics}
\usepackage{color,ulem}
\allowdisplaybreaks

\usepackage{tikz}
\usetikzlibrary{trees}
\usetikzlibrary{decorations.pathmorphing}
\usetikzlibrary{decorations.markings}
\usetikzlibrary{decorations, decorations.markings, decorations.pathmorphing, 
arrows, graphs, shapes.geometric, snakes}
\usetikzlibrary{arrows}

\tikzset{
photon/.style={decorate, decoration={snake}, draw=red},
dark/.style={draw=gray, postaction={decorate},
        decoration={markings,mark=at position .55 with {\arrow[draw=gray]{>}}}
        },
antidark/.style={draw=gray, postaction={decorate},
        decoration={markings,mark=at position .55 with {\arrow[draw=gray]{<}}}
        },
electron/.style={draw=violet, postaction={decorate},
        decoration={markings,mark=at position .55 with {\arrow[draw=violet]{>}}}
        },
neutrino/.style={draw,color=violet,thick, postaction={decorate} },
neutrinolight/.style={draw=blue, postaction={decorate} },
quark/.style={draw=blue, postaction={decorate},
        decoration={markings,mark=at position .55 with {\arrow[draw=blue]{>}}}},
antiquark/.style={draw=blue, postaction={decorate},
        decoration={markings,mark=at position .55 with {\arrow[draw=blue]{<}}}},
heavyquark/.style={draw=purple, postaction={decorate},
        decoration={markings,mark=at position .55 with {\arrow[draw=purple]{>}}}
        },
antiheavyquark/.style={draw=purple, postaction={decorate},
        decoration={markings,mark=at position .55 with {\arrow[draw=purple]{<}}}
        },
        gluon/.style={decorate, draw=or,
        decoration={coil,amplitude=2pt, segment length=3pt}},
gluon/.style={decorate, draw=or,
        decoration={coil,amplitude=2pt, segment length=3pt}},
ZZ/.style={decorate, decoration={snake,amplitude=1.5pt, segment length=5pt}, 
draw=greeen}, left }

\definecolor{greeen}{rgb}{0.03,0.84,0.13}
\definecolor{test}{rgb}{0.03,0.74,0.33}
\definecolor{viol}{rgb}{0.44,0,0.94}
\definecolor{or}{rgb}{0.95,0.65,0}

\newcommand{\red}{\textcolor{red}}
\newcommand{\blue}{\textcolor{blue}}
\newcommand{\green}{\textcolor{green}}

\def\lsim{\mathrel{\vcenter{\hbox{$<$}\nointerlineskip\hbox{$\sim$}}}}
\def\gsim{\mathrel{\vcenter{\hbox{$>$}\nointerlineskip\hbox{$\sim$}}}}

\begin{document}
%%%%%%%%%%%%%%%%%%%%%%%%%%%%%%%%%%%%%%%
\preprint{UMD-PP-018-03}

\title{Leptonic $CP$ Violation and Proton Decay in $SUSY$ $SO(10)$ }

\author{Rabindra N. Mohapatra}\author{and Matt Severson}
\affiliation{Maryland Center for Fundamental Physics and Department of Physics,
University of Maryland, College Park, Maryland 20742, USA}

%\date{\today}
%%%%%%%%%%%%%%%%%%%%%%%%%%%%%%%%%%%%%%%
\abstract{  We study the correlation between proton lifetime and
leptonic CP violation in a class of renormalizable supersymmetric SO(10) grand
unified theories (GUTs) with {\bf 10}, {\bf 126} and {\bf 120} Higgs
fields, which provides a unified description of all fermion masses and
possibly resolution of the strong CP problem. This specific model is
unique in that it can readily be compatible with current proton
lifetime limits for a supersymmetry (SUSY) breaking scale as low as 5
TeV due to the presence of a specific Yukawa texture. Our
investigation here reveals that proton partial lifetimes predicted by
this class of models will be tested by forthcoming proton decay
searches; furthermore, a discovery of leptonic $CP$ violation in
neutrino oscillations would also lead to substantial reduction of the
parameter space of the model.}
%\end{abstract}
%%%%%%%%%%%%%%%%%%%%%%%%%%%%%%%%%%%%%%%
\maketitle
%%%%%%%%%%%%%%%%%%%%%%%%%%%%%%%%%%%%%%%
\section{Introduction}\label{sec:1} One of the major challenges for
beyond the standard model physics is finding a unified understanding
of all fermion masses including neutrinos. The origin and pattern of
neutrino masses and mixings observed in various oscillation
experiments over the past two decades not only reveals an absolute
scale for neutrino masses substantially smaller than other fermions
but also a leptonic mixing pattern quite different from the quarks.
%There are essentially two theoretical issues in neutrino mass
%physics:\;(i) why are the neutrino masses so much smaller than quark
%and charged lepton masses, and (ii) what are the origins of the
%mixing patterns of the quark and lepton sector, and why are they so
%different from each other? 
While the seesaw mechanism~\cite{seesaw} is supposed to provide a
simple path to the resolution of the scale problem, the problem of
mixing is more complicated, and some believe that could be signal of
new leptonic symmetries~\cite{king}. In particular as experiments make
progress toward answering all remaining questions in neutrino mass
physics, such as the type of mass hierarchy, presence of CP violation,
and Dirac vs Majorana nature, it is important to sharpen the
predictions of various theories of neutrino masses, so that a clearer
picture of the direction of physics beyond the standard model can be
determined.

Grand unified theories~\cite{raby} are a surprisingly appropriate
setting for the seesaw mechanism for a number of reasons: (i) SO(10)
theories \cite{so10} have $B-L$ as a subgroup, which is naturally
associated with the seesaw scale; (ii) the matter spinor of SO(10)
automatically contains the right-handed neutrino field as a symmetry
partner to the quarks and leptons of the standard model; (iii) SO(10)
theories relate the Dirac mass of neutrinos to the up-like quark
masses, leading to a seesaw scale near the GUT scale, which connects
it to the scale of coupling unification as well. 

In this paper, we discuss one class of supersymmetric SO(10) grand
unified models including neutrinos and based on renormalizable Yukawa
interactions \cite{BM,other,120,120other}. The constraints of grand
unification for this particular class are strong enough to make the
model quite predictive in the fermion sector, due to fewer
parameters in the Yukawa couplings~\cite{BM}. For this class of
models, we continue the exploration from previous work \cite{DMM,matt}
with the goal of extracting predictions for CP violation in neutrino
oscillation and corresponding predictions for proton decay that can be
used to test them. 

As is well known, in SO(10) models, the fermions of each generation
are assumed to reside in the {\bf 16}-dimensional spinor
representation, and there are three kinds of Higgs field
representations that can give renormalizable Yukawa couplings to
generate fermion masses: ${\bf 10} \equiv {\rm H}$, $\overline{\bf
126}\equiv \bar{\Delta}$ and ${\bf 120}\equiv \Sigma$. The
$\bar\Delta$-field is also used also to break the $B-L$ gauge symmetry
of the model, and since it breaks $B-L$ by two units, this preserves
discrete R-parity symmetry~\cite{RPV}, leading to a [naturally stable
lightest SUSY partner. Within this general framework,
two kinds of renormalizable models have been discussed in the
literature:\;(I) One that uses only ${\bf 10}+\overline{\bf 126}$
Higgs superfields to give mass to the fermions which reside in the
{\bf 16}-dim spinor representations of SO(10)~\cite{BM,other} and (II)
another that uses all three representations, {\it i.e.} ${\bf
10}+\overline{\bf 126}+{\bf 120}$ \cite{120,120other}. The GUT
symmetry is broken by a subset of {\bf 210} ($\Phi$), {\bf 45} (A),
{\bf 54} (S) fields, which do not affect the fermion predictions. The
first class of models with only {\bf 210} to break GUT symmetry was
suggested early on as a minimal SUSY SO(10) class with coupling
unification \cite{aulakh}. Both these classes provide a relatively
economical way to fit all fermion masses and mixings, including
neutrinos as well as quarks and charged leptons, and lead to
predictions such as (i) normal hierarchy for neutrinos, (ii)
$\theta_{13}$ in agreement with observation \cite{theta13}, and (iii)
CP violating phases for leptons, using a combination of both type I
and type II seesaw~\cite{type2}. 

What is in some sense amazing about this class of models is that
despite the fact the quarks and leptons are unified, the diverse
mixing patterns among them emerge without fine tuning. Especially in
models that use ${\bf 10}+ \overline{\bf 126}$ fields for mass
generation and where type II seesaw is assumed to dominate, an elegant
explanation of the correct mixings emerges from the dynamical reason
that in simple grand unified theories, bottom quark and tau lepton
masses, become very close to each other when run up to the GUT
scale~\cite{BSV}. This feature plays an important role in generating
correct neutrino mixings in the general renormalizable SO(10) models
as well.

Successes of these models in the fermion sector, as well as their
economy, have led to further scrutiny of other model predictions
accessible by experiment. Primary among such tests is the classic
prediction of proton decay in grand unified theories and the
corresponding question of proton lifetime \cite{pdk}. An advantage of
renormalizable SUSY SO(10) models is that the same parameters that go
into obtaining fermion mass fits also contribute to proton decay so
that neutrino mixing angles and CP violating phases are expected to be
correlated with proton lifetime, making it possible to test the models
using measurements of those parameters together with those of proton
lifetime.

This has been investigated in the framework of models with only ${\bf
10}+\overline{\bf 126}$ Higgs superfields contributing to fermion
mass, and there seems to be tension with current data
\cite{pdecay,babu} unless the SUSY breaking scale is made high to
suppress the $d=5$ contribution to p-decay amplitude. In fact it has
been shown in ~\cite{babu} that the current limit on the mode $p\to
K^+\nu$ implies a lower limit on the SUSY breaking scale of $M_{susy}\geq
238$ TeV, which is much higher than the conventionally assumed value.
Indeed, if any evidence for supersymmetry appears at the LHC energies,
the ${\bf 10}+\overline{\bf 126}$ will be ruled out.

It was suggested in \cite{DMM} that with a choice of specific Yukawa
textures, possible only in models with ${\bf 10}+\overline{\bf
126}+{\bf 120}$ due to fermion fit constraints, one can accommodate
the proton lifetime bounds while keeping $M_{susy}$ in the low TeV
range. Detailed investigation of proton decay predictions of this
model with $M_{susy}\sim 5$ TeV was carried out in ~\cite{matt}, and
the results were found to be promising. The Dirac CP violating phase
however was found to be $\delta_{CP}\sim -7^\circ$ for the type II
case, while the type I fit yielded a value of $\delta_{CP}\sim
-46^\circ$ \cite{matt}.  Meanwhile, the NOVA and T2K results are
providing hints of a larger $\delta_{CP}$~\cite{Nova}.

In this paper, we discuss three new points beyond the analysis in
Ref.~\cite{matt}: (i) we explore a much wider domain of parameter
space to see if the model can accommodate larger CP phase. We answer
this question in the affirmative for the type I seesaw case and in the
negative for type II case. (ii) We show how the CP phase predictions
are correlated with predictions for proton lifetime. We believe that
this result should be interesting since it is a primary goal of
several planned experiments~\cite{hyperK,DUNE,T2K,JUNO} to search for
both proton decay and leptonic CP violation, which will then put this
interesting class of models ``under the microscope." (iii) We also
discuss how models of this type can provide a solution to the
strong CP problem in a grand unified theory context without the need
for an axion.  On this point, we only show the result at the tree
level and leave detailed investigation of any loop effects to a
separate investigation.

%Incidentally, we note that SO(10) models of this class but
%without supersymmetry have also been shown to lead to successful results
%for the neutrino sector~\cite{nonsusy}; however, in such models, proton
%decay arises from gauge boson exchanges and depends only on the
%unification scale, so the deep and more direct connection to neutrino
%mixing patterns seen in models with SUSY does not exist.

Our calculation procedure is as follows: using the fermion mass sum
rules predicted by these models, we calculate the GUT scale masses and
mixings for the quarks and leptons and, through minimization of
$\sum\chi^2$, attempt to locate fits to fermion sector measurements
with CP phase outputs spanning the range of possible values. We create
plots from the output data to observe which CP phase values are most
favorable for the model. We then use the fermion fit data and varied
values for the parameters of the proton decay operators (both LLLL and
RRRR modes) to determine the $p \to K^+\bar\nu$ partial lifetime
corresponding to each value for the CP phase, which illuminates
whether phase values favored by the fermion sector are also consistent
with proton lifetime constraints. We take gaugino masses to be 300 GeV
and squark masses to be 5 TeV, although current data does not require
us to use such high values. Predictions for lower SSB parameters can
be easily obtained using scaling. The results we report can be used to
test the model with the forthcoming experiments \cite{hyperK,DUNE,T2K,JUNO}.

This paper is organized as follows: in Sec.\,\ref{model} we review the
salient features of the model; Sec.\,\ref{cp} is devoted to a
discussion of how CP violation is introduced into this model, and how
our approach hints at a new way to solve the strong CP problem; in
Sec.\,\ref{proton} we discuss proton decay operators in the model;
Sec.\,\ref{fit} is devoted to fermion mass fitting and CP phase
predictions in the model; and in Sec.\,\ref{pfit}, we present our
predictions for proton lifetime and its correlation with the Dirac CP
phase.  Sec.\,\ref{conc} gives discussion and conclusion.
%%%%%%%%%%%%%%%%%%%%%%%%%%%%%%%%%%%%%%%
%%%%%%%%%%%%%%%%%%%%%%%%%%%%%%%%%%%%%%%

\section{Details of the Model} \label{model} The supersymmetric
$SO(10)$ model we consider has \textbf{10}-, $\overline{\bf{126}}$-,
and \textbf{120}-dimensional Higgs fields with renormalizable Yukawa
couplings contributing to fermion masses. The fields are named here as
H, $\overline{\Delta}$, and $\Sigma$, respectively. The relevant
Yukawa superpotential terms are
\begin{equation}
  W_Y ~\ni ~h_{ij} \Psi_i \Psi_j {\rm H} + f_{ij} \Psi_i \Psi_j
  \overline{\Delta} + g_{ij} \Psi_i \Psi_j \Sigma,
\label{eq:W}
\end{equation}
where $\Psi_i$ is the \textbf{16}-dimensional matter spinor containing
superfields of all the SM fermions (of one generation) plus the
right-handed neutrino, and $i$ is the generation index. Additional
multiplets such as ${\bf{126}}(\Delta)$, ${\bf 210}(\Phi)$ and ${\bf
54}$(S) are needed to break the GUT symmetry and maintain
supersymmetry below the GUT scale down to TeV scale. They contain in
their decompositions color triplets that contribute to proton decay,
as well as $SU(2)$ doublets that contribute to the effective MSSM
Higgs doublets, but they do not contribute to fermion masses.

All the doublets contained in the decompositions of H,
$\overline{\Delta}$, $\Sigma$, and the other GUT-scale Higgs fields
mix with each other in the mass matrix $\mathcal{M_D}$ defined by
$\varphi_u^T \mathcal{M_D} \,\varphi_d$, which is diagonalized by the
bi-unitary rotation ${\cal U\, M_D V}^T$. One linear combination each
of the $\varphi_{u,d}$ fields becomes the nearly massless MSSM Higgs
doublet $H_{u,d}$, while the other combinations remain heavy.  For
$H_{u,d}$ to remain light, the condition $\det\mathcal{M_D}\sim0$ is
necessary. 

%The MSSM Higgs doublets transform as $\left(\,{\bf 1}, {\bf 2},
%\pm\frac{1}{2} \,\right)$ under the standard model gauge group,
%$SU(3)_c\times SU(2)_L\times U(1)_Y$; they emerge as linear
%combinations of doublets with same representation content contained in
%the decompositions of H, $\overline{\Delta}$, $\Sigma$, and the other
%GUT-scale Higgs fields. All the doublets mix with each other in a
%single mass matrix $\mathcal{M_D}$ defined by $\varphi_u^T
%\mathcal{M_D} \,\varphi_d$, in the basis where $\varphi_u = ({\rm
%H}_u, \Sigma^1_u, \Sigma^{15}_u, \bar{\Delta}_u, \Delta_u, \Phi_u)$,
%and similar for $\varphi_d$. When this mass matrix is diagonalized by
%the bi-unitary rotation ${\cal U\, M_D V}^T$, one linear combination
%each of the $\varphi_{u,d}$ fields becomes the nearly massless MSSM
%Higgs doublet $H_{u,d}$, while the other combinations remain heavy.
%For $H_{u,d}$ to remain light, the condition $\det\mathcal{M_D}\sim0$
%is necessary. 
%This condition can be interpreted as the fixing of one parameter in
%the matrix to an $\mathcal{O}(1)$ value; the parameter conventionally
%chosen is the mass $M_H$ of the \textbf{10} field. 

%\subsection{Fermion mass formulae}
The full details of the mass matrices in terms of GUT-scale vevs and
parameters are given for this type of model in \cite{aulgarg}. The
resulting effective Dirac fermion mass matrices can be written
as~\cite{120}
\begin{align}
  {\cal M}_u &= \tilde{h}+r_2 \tilde{f}+r_3\tilde{g} \nonumber \\ 
  {\cal M}_d &= \frac{r_1}{\tan\beta}
  (\tilde{h}+\tilde{f}+\tilde{g}) \nonumber \\
  {\cal M}_e &= \frac{r_1}{\tan\beta}
  (\tilde{h}-3\tilde{f}+c_e\tilde{g}) \nonumber \\
  {\cal M}_{\nu_D} &= \tilde{h} - 3 r_2 \tilde{f} + c_\nu \tilde{g},
  \label{eq:mass}
\end{align}
with $\tan \beta = v_u / v_d$, where $v_{u,d}$ are vevs of the MSSM
fields $H_{u,d}$. For $\lambda = h,f,g$, the couplings
$\tilde{\lambda}_{ij}$ are related to $\lambda_{ij}$ from
eq.\,(\ref{eq:W}) by \\ \cite{DMM} 
\begin{equation} 
  \tilde h \equiv {\cal V}_{11} h\, v_u; \quad \tilde f \equiv
  \frac{{\cal U}_{14} f v_u}{r_1 \sqrt{3}}; \quad \tilde g \equiv
  \frac{{\cal U}_{12} + {\cal U}_{13}/\sqrt{3}}{r_1} g\, v_u,
  \label{eq:tildes}
\end{equation} \\
where $1 / \tan \beta$ takes $v_u \rightarrow v_d$ for down-type
fields; the vev ratios $r_i$ and $c_\ell$ are given by \\
\begin{gather}
  r_1 \equiv \frac{{\cal U}_{11}}{{\cal V}_{11}}; \quad
  r_2 \equiv r_1 \frac{{\cal V}_{15}}{{\cal U}_{14}}; \quad r_3 \equiv
  r_1 \frac{{\cal V}_{12} - {\cal V}_{13}/\sqrt{3}}{{\cal U}_{12} +
  {\cal U}_{13}/\sqrt{3}}; \nonumber \\ c_e \equiv \frac{{\cal U}_{12}
  - {\cal U}_{13} \sqrt{3}}{{\cal U}_{12} + {\cal U}_{13}/\sqrt{3}};
  \quad c_\nu \equiv r_1 \frac{{\cal V}_{12} + {\cal V}_{13}
  \sqrt{3}}{{\cal U}_{12} + {\cal U}_{13}/\sqrt{3}},
\end{gather} \vspace{1mm}

The full neutrino mass matrix is determined by both Majorana mass
terms in the superpotential and the Dirac mass contribution given in
eq.\,(\ref{eq:mass}). The light masses can be generally given by a
combination of the type-I and type-II seesaw mechanisms, involving the
vevs of both left- and right-handed Majorana terms:
\begin{equation}
  {\cal M}_\nu = v_L f - {\cal M}_{\nu_D}\left(v_R f\right)^{-1}\left(
  {\cal M}_{\nu_D}\right)^T,
\label{eq:neu}
\end{equation}
where $v_{L,R}$ are the vevs of the SM-triplet $\overline{\Delta}_{L}$
and singlet $\overline{\Delta}_{R}$, respectively, in $\overline{{\bf
126}}$. We will separately consider cases of type-II ($v_L$ term) and
type-I ($1/v_R$ term) dominance. Note that the presence of the $f$
coupling in both terms intimately connects the neutrino mass matrix
properties to those of the charged sector matrices, making the model
quite predictive. Also note we will consider only normal mass
hierarchy in this analysis, which arises naturally in this class of
models.
\section{CP Violation, Strong CP, and Yukawa Texture} \label{cp} In
our phenomenological analysis, we will assume that $\tilde{h}$,
$\tilde{f}$ are real and $\tilde{g}$ is imaginary. Note that in the
original model of this type with {\bf 120}, we could always choose
$h_{ii}$ to be real, but to make $f_{ij}$ real and $g_{ij}$ imaginary,
we have to rely on some extra assumptions.  We could view this choice
simply as a way to fit observations; however, we argue below that this
choice for the mass matrix parameters could arise from an underlying
symmetry of the theory. We discuss two potential paths to obtaining CP
violation naturally, and find that one of the ways could provide a
path to resolution of the strong CP problem without the need for an
axion. We show that in this case, the $\theta$-parameter is zero only
at the tree level due to Hermiticity of the quark mass matrices;
we defer a detailed discussion of this issue to a latter paper. 

\subsection{CP Violation and Potential Solution to Strong CP} In this
subsection we discuss the first approach to CP violation in the model.
First we assume that theory is invariant under CP transformation of
the fields as seen in Table \ref{table:cp}, so that CP is
spontaneously broken. 

\begin{table}\centering
\begin{tabular}{|c||c|}\hline
Field & CP transform\\\hline
$\Psi(16)$ & $\Psi^*(16)$\\
${\rm H}(10)$ & ${\rm H}^*(10)$\\
$\bar{\Delta}(\overline{126})$ & $\bar{\Delta}^*(\overline{126})$\\
${\Delta}({126})$ & $-{\Delta}^*({126})$\\
$\Sigma(120)$ & $-\Sigma^*(120)$\\
${\rm A}(45)$&${\rm -A}^*(45)$\\
${\rm S}(54)$&${\rm S}^*(54)$\\
${\rm X}(1)$&${\rm X}^*(1)$\\\hline
\end{tabular}
%\vspace{2mm}
\caption{CP transformations of the superfields of the theory in the
first approach.} \label{table:cp}
\end{table}
\noindent We supplement this transformation with a $Z_2$ symmetry
under which only the $\Delta(126)$ field is odd while all other fields
are even. Invariance under this symmetry implies that Yukawa couplings
$h,f$ are real and $g$ is imaginary. After GUT symmetry breaking, one
linear combination of the SM Higgs doublet fields remains massless,
and it contains no complex parameters. To see this, first note that we
do not have a {\bf 210} Higgs superfield in the theory. The
superpotential for the symmetry breaking sector can then be written
as:
\begin{gather}
{\cal W}~=~\sum_{\varphi} M_\varphi \,\varphi^2 +
 \lambda_1 {\rm X}\, ({\rm A}^2-M^2_U) +
\lambda_2\, \Sigma\, {\rm A\, H} + \nonumber \\
\frac{\lambda_3}{\Lambda}\, \bar{\Delta}\, {\rm A^2 H} +  \lambda_5\,
{\rm S\,A\,A} + \lambda_6\, {\rm S\,H\,H} + \frac{1}{\Lambda}(\Delta
\bar{\Delta})^2 \nonumber \\[3mm] +\lambda_7\,{\rm S} \Delta\Delta +
\lambda_8\, {\rm S} \bar{\Delta} \bar{\Delta}
\end{gather} \label{eq:cpW}
\noindent with $\varphi = {\rm H}, \Sigma, {\rm A, S}$.

Note that all superpotential parameters are real, and therefore we
expect all resulting vevs to be real; furthermore, no new phases are
expected to ``sneak" in during the process of deriving the low-energy
effective Lagrangian. After breaking the SM symmetry by the vevs of
the Higgs fields (expected to be real), the resulting quark mass
matrices are Hermitian at tree level and therefore have ${\rm
Arg\,(Det}~M_q) = 0$, which gives no contribution to the strong CP
phase $\theta$ at tree level; additionally, due to reality of all the
parameters in the superpotential, all heavy colored fermions also have
real masses, so there is no contribution to the tree-level from them
either. Thus at tree level $\theta = 0$ at the GUT scale
\cite{strongCP}, and the model has the potential to give a solution
for the strong CP problem without the need for an axion.

The loop effects (or RGE effects) on $\theta$ are currently under
study and are beyond the scope of this paper; for now we simply use
this result to justify our choice for the form of the Yukawa
couplings.
%We do not claim that this is a full solution to the strong CP
%problem- but simply a possible pathway to that.

\subsection{CP Violation from Vacuum CP Phase} A second approach to CP
violation is to assume that the theory is again invariant under
$CP\times Z_2$, but instead all fields transform as $\Phi \to \Phi^*$
under CP. This implies that the matrices $h$ and $f$ are real and
symmetric matrices and $g$ is real and anti-symmetric. 
%We further assume that there are the following GUT breaking Higgs
%multiplets {\bf 45}, {\bf 210} and {\bf 54}. 
In this case, the {\bf 45} field A responsible for GUT symmetry
breaking gets an {\it imaginary vev} via the F-term of a singlet X,
through the superpotential term $\lambda\, {\rm X}\, ({\rm
A}^2+M_U^2)$ (note the change of sign within the bracket compared to
the corresponding term in eq.\,(\ref{eq:cpW})).  This imaginary vev
leads to an imaginary mixing term between the doublets H$_{u,d}$ and
$\Sigma_{u,d}$ from {\bf 10} and {\bf 120}, respectively. Then, by
redefining the doublets in {\bf 120} by $\Sigma_{u,d} \to
i\Sigma_{u,d}$ and keeping all other fields as they are, the effective
Yukawa coupling of {\bf 120} doublets to matter fields becomes
imaginary, whereas all other Yukawa couplings remain real. Below the
GUT scale, the two linear combinations that become the MSSM doublets
have all coefficients real since their mass matrices are real. In the
language of mixings \,${\cal U}_{ij}$, one sees that \, ${\cal U}_{12,
13}$ and  ${\cal V}_{12, 13}$ are imaginary and all other elements
real. Whether this case is also a potential solution to strong CP is
not clear due to the color triplet fields having also
imaginary coefficients after redefinition.%Additional
%details on how this comes about, as well as which conditions must be
%satisfied by parameters involving the Higgs fields, is given in
%Appendix A.

\subsection{Parameter counting} With either of the above choices of
symmetry, and in the absence of any texture assumptions, the mass
matrices for the fermions are now Hermitian and contain a total of 12
parameters from Yukawa couplings; seven additional parameters from
doublet vevs, including $v_L$ in the type-II contribution, bring the
total number of parameters to 18; finally, three more parameters arise
once we account for threshold effects, bringing the total to 21.

To reduce the number of parameters, and also to ameliorate the proton
decay problem of these models, we consider the reduced Yukawa texture
proposed and analyzed in \cite{DMM} and further studied in
\cite{matt}. The proposed~\cite{DMM} texture has the form \vspace{1mm}
\begin{gather}
  \tilde{h} = \left(\begin{array}{ccc} 
    0 & & \\
    & 0 & \\
    & & M \end{array}\right), \qquad
  \tilde{f} = \left(\begin{array}{ccc}
    \sim 0 & \sim 0 &\tilde  f_{13}\\
    \sim 0 &\tilde  f_{22} &\tilde f_{23}\\
  \tilde  f_{13} &\tilde f_{23} &\tilde f_{33}\end{array}\right), \nonumber \\[4mm]
  \tilde{g} = i \left(\begin{array}{ccc} 
    0 &\tilde g_{12} & \tilde g_{13} \\
    -\tilde g_{12} & 0 & \tilde g_{23} \\
    -\tilde g_{13} & -\tilde g_{23} & 0 \end{array}\right) \label{eq:y},
\end{gather}
\\ where we note that $\tilde{h}$ is an explicitly
rank-1 matrix, with $M \sim m_t$; thus, at first order, the
\textbf{10} Higgs contributes to the third generation masses and
nothing more.

The above matrices have 10 parameters (with $f_{11}$ and $f_{12}$
small but nonzero and not shown in the above equation), so taken in
combination with $v_L$ and the vev mixing ratios $r_i$ and $c_\ell$,
the model has a total of 16 parameters.  Correspondingly, there are 18
measured parameters associated with the physical fermions, with limits
on the Dirac CP phase in the PMNS matrix beginning to materialize.
Hence, we would like to predict 19 observables in the fermion sector.
The model is surprisingly close to fitting all 19 values with its 16
parameters, but ultimately the inclusion of SUSY threshold
corrections, which provides three additional parameters (expected to be
significant for large $\tan \beta$), are needed to 
get a good fit~\cite{matt2}; they provide the model with the
three additional degrees of freedom needed to accommodate all 19
observables, without tension from squark and Higgs mass constraints in
the MSSM sectors \cite{matt2} \cite{matt}.

Note that the magnitude constraints on the $f_{ij}$ and
$g_{ij}$ required by the texture ensatz, as well as phenomenological
constraints on the threshold corrections \cite{matt2}, prevent a true
$\sum\chi^2 = 0$ fit to the data despite the sufficient naive match
between input and output counts.

\section{Proton Decay in the Model} \label{proton} The
baryon-number violating operators leading to proton decay with which
we are concerned in this paper follow from the presence of $SU(3)$
color-triplets $\left(\,{\bf 3},{\bf 1},-\frac{1}{3} \,\right)$ +
c.c\, in the decompositions of the GUT Higgs fields; these triplets
mix with each other in a mass matrix ${\cal M_T}$, which is
diagonalized by the bi-unitary rotation ${\cal X\, M_D Y}^T$. Two
exotic types of triplets also lead to $B$- or $L$-violating vertices,
$\left(\,{\bf 3},{\bf 1}, -\frac{4}{3} \right)$ + c.c, which interact
with two up-type or two down-type RH singlet fermions, and
$\left(\,{\bf 3},{\bf 3}, -\frac{1}{3} \right)$ + c.c, with a pair of
LH doublets. The exotic triplets mix in their own respective
(2$\times$2) matrices. All such triplets are expected to be heavy with
masses near the GUT scale.

\begin{figure}[t]
\begin{center}
  \includegraphics{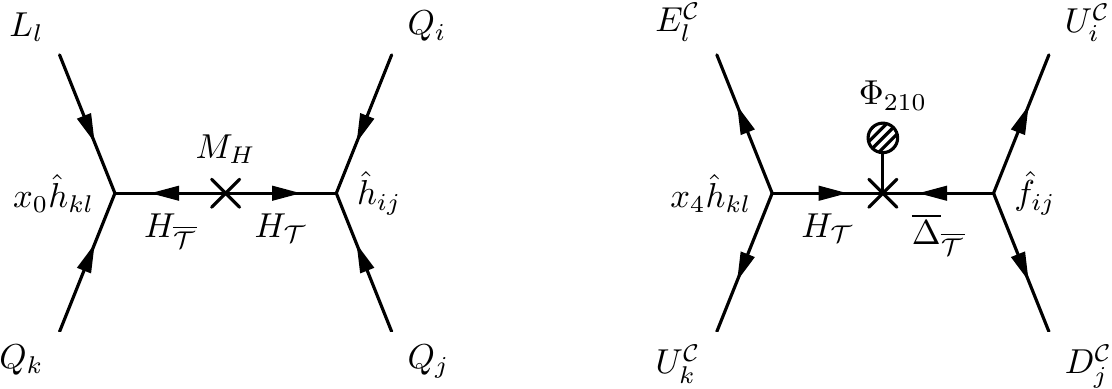}
    \caption{\footnotesize Examples of superfield diagrams that lead to
    proton decay in this model. The hats on the couplings indicate mass
    basis, and the parameters $x_i$ contain the triplet mixing
    information unique to the specific pairing of couplings present in
    each diagram (see below).} %\vspace{-8mm}
    \label{fig:susyfd}
\end{center}
\end{figure}

Exchange of conjugate pairs of any of these triplets leads to
 operators that change two quarks into an anti-quark and a lepton;
such operators are numerically dominant over the corresponding
exchange of the scalar superpartners of these triplets, and
may be so over gauge boson exchange as well. Figure \ref{fig:susyfd}
shows Feynman diagrams for two examples of the operators in question.

The corresponding $d=5$ effective superpotential is
\begin{equation}
  {\cal W}_{\Delta B = 1} = \frac{\epsilon_{abc}}{M_\mathcal{T}}
  \left( \widehat{C}^L_{ijkl} Q^a_i Q^b_j Q^c_k L_l +
  \widehat{C}^R_{[ijk]l} U^{\mathcal{C}\,a}_i D^{\mathcal{C}\,b}_j
  U^{\mathcal{C}\,c}_k E^\mathcal{C}_l \right),
\label{eq:effW}
\end{equation}
where $i,j,k,l = 1,2,3$ are the generation indices and $a,b,c = 1,2,3$
are the color indices. This potential has $\Delta L = 1$ in addition
to $\Delta B = 1$ and so also has $\Delta(B - L) = 0$. $M_\mathcal{T}
\sim M_U$ is a generic GUT-scale mass for the triplets.
The left-handed term expands to
\begin{equation}
  {\cal W}_{\Delta B = 1} \ni \frac{\epsilon_{abc}}{M_\mathcal{T}}
  \left( \widehat{C}^L_{\{[ij\}k]l} U^a_i D^b_j U^c_k E_l -
  \widehat{C}^L_{\{i[j\}k]l} U^a_i D^b_j D^c_k \mathcal{N}_l \right),
\label{eq:effWL}
\end{equation}
where $\mathcal{N}$ is the left-handed neutrino superfield. The
anti-symmetrizing of indices in the $C_R$ operator from
(\ref{eq:effW}) and here in the $C_L$ reflects the non-vanishing
contribution from the contraction of the color indices. The
symmetrization in $i,j$ in the $C_L$ comes as a result of the doublet
contractions. 

%\newpage
The effective operator coefficients $C_{ijkl}$ are of the form
\begin{align}
  C^L_{ijkl} &= \sum_a \left( {\cal X}_{a1} h + {\cal X}_{a4} f +
  \sqrt{2}\, {\cal X}_{a3}\, g \right)_{ij} 
  \left( {\cal Y}_{a1} h + {\cal Y}_{a5} f\, \right)_{kl} ~+ ~
  {\rm exotic~terms}; \\
  C^R_{ijkl} &= \sum_a \left( {\cal X}_{a1} h - {\cal X}_{a4} f +
  \sqrt{2}\, {\cal X}_{a2}\, g
  \right)_{ij} \left[ {\cal Y}_{a1} h - \left( {\cal Y}_{a5} - \sqrt{2}
  {\cal Y}_{a6} \right) f 
  +  \sqrt{2}  \left( {\cal Y}_{a3} - {\cal Y}_{a2} \right) g
  \right]_{kl} \nonumber \\ 
  &~ + {\rm exotic~terms}
  \label{eq:Cexact}
\end{align}
where the matrices ${\cal X}$ and ${\cal Y}$ are those that
diagonalize the triplet mass matrix; hence elements of those matrices
will be between 0 and 1. These elements are algebraic combinations of
the various GUT scale masses and couplings; again the details of their
properties can be seen in \cite{aulgarg}. Nearly all such parameters
are essentially unconstrained and can be taken as arbitrary for the
purposes of our analysis. The one exception here is the product ${\cal
X}_{a1} {\cal Y}_{a1} \sim M_H$, which is fixed by the tuning
condition for $M_\mathcal{D}$ discussed above. As a result, the value
of this product needs to be generally ${\cal O}(1)$.

%\newpage
Due to the largely unconstrained nature of the remaining parameters,
it is typical to re-cast the $C_{ijkl}$ coefficients in the following
parametrization:
\begin{align}
  C^L_{ijkl} &= x_0 h_{ij} h_{kl} + x_1 f_{ij} f_{kl} -
  x_3 h_{ij} f_{kl} - x_4 f_{ij} h_{kl} + y_5 f_{ij} g_{kl} +
  y_7 h_{ij} g_{kl} \nonumber \\
  & + y_9 g_{ik} f_{jl} + y_{10} g_{ik} g_{jl}. \nonumber \\
  C^R_{ijkl} &= x_0 h_{ij} h_{kl} + x_1 f_{ij} f_{kl} +
  x_2 g_{ij} g_{kl} + x_3 h_{ij} f_{kl} +
  x_4 f_{ij} h_{kl} + x_5 f_{ij} g_{kl} \nonumber \\
  & + x_6 g_{ij} f_{kl} + x_7 h_{ij} g_{kl} +
  x_8 g_{ij} h_{kl} + x_9 f_{il} g_{jk} +
  x_{10} g_{il} g_{jk}, 
  \label{eq:Cs}
\end{align}
where $x_0 \equiv {\cal X}_{a1} {\cal Y}_{a1} \sim 1$. Note that
several identifications have already been made here: $y_{0,1} =
x_{0,1}$ and $y_{3,4} = -x_{3,4}$; the would-be parameters $y_{2,6,8}
= 0$. The parameters $x_{9,10}$ and $y_{9,10}$ correspond to the
exotic triplets; the indices of those terms are connected in unique
ways as a result of the distinct contractions of fields.

The couplings $h,f,g$ as written correspond to matter fields in the
flavor basis and undergo unitary rotations in the change to mass
basis, as indicated by the hats on $\widehat{C}^{L,R}$ in
eqs.\,(\ref{eq:effW}) and (\ref{eq:effWL}) above. For more general
Yukawa textures, this change of basis does not alter the prescription
significantly, and so the corresponding rotations are often overlooked
in similar analyses; for this work however the rank-1 nature of the
{\bf 10} coupling results in an increased sensitivity to the basis
change, and so it cannot be safely ignored. I will return to the
details of the pertinent rotations in basis after introducing the
contributing operators.
\begin{figure}[t]
\begin{center}
  \includegraphics{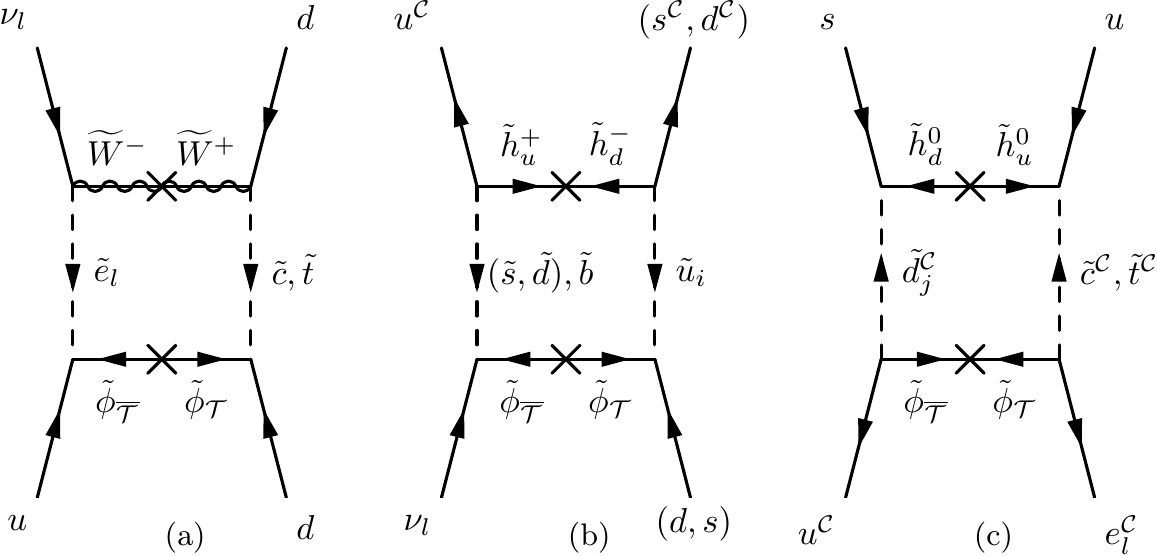}
  \caption{\footnotesize Examples of dressed diagrams leading to proton
  decay in the model. Diagram (a) shows a contribution to $p \rightarrow
  \pi^+ \bar{\nu}_l$; integrating out the triplets gives an effective
  operator of type $C^{L}udue$.  Diagram (b) shows a $C^{L}udd\nu$-type
  operator contributing to $K^+ \bar{\nu}_l$.  Diagram (c) shows a
  $C^{R} u^\mathcal{C} d^\mathcal{C} u^\mathcal{C} e^\mathcal{C}$-type
  operator contributing to $K^0 e_l^+$, for $l=1,2$. Note where more
  than one field is listed, each choice gives a separate contributing
  channel, except for the dependent exchange of $(s \leftrightarrow d)$
  in (b).} %\vspace{2mm}
  \label{fig:compfd}
\end{center}
\end{figure}

\subsection{Dressed Operators}
The pertinent superfield operators are realized at leading order
through conjugate pairs of Higgsino triplet mediators, and the
outgoing squarks and sleptons interact with gauginos or (SUSY)
Higgsinos to produce the corresponding outgoing fermions, as seen in
Figure \ref{fig:compfd}. The resulting $d=6$ effective operators have
the four-fermion form needed to enable proton decay.

Depending on the sfermions present, diagrams may in principle be
dressed with gluinos, Winos, Binos, or Higgsinos.\footnote{We give
this discussion in terms of $\widetilde{B}$, $\widetilde{W}^0$, and
$\tilde{h}_{u,d}^{\pm,0}$, rather than $\widetilde{\chi}^\pm$ and
$\widetilde{\chi}^0_i$, because $(a)$ we assume universal mass
spectrum for superpartners to satisfy FCNC constraints, meaning the
mass and flavor eigenstates coincide for the gauge bosons, and $(b)$
chargino and neutralino masses differ from $M_{SUSY}$ by
$\mathcal{O}(1)$ factors as long as gaugino soft masses are relatively
small compared to $M_{SUSY}$. These assumptions simplify
computations with little affect on the numerical factors.} Gluino,
Bino, and $\widetilde W^0$ operators can contribute to only the $p
\rightarrow K^+ \bar{\nu}_l$ mode, through the $UDS\mathcal{N}$
operator (left-handed only), due to generation diagonality of the
interactions and generation anti-symmetry of the quarks in the
$C_{ijkl}$. Furthermore, those contributions from gluino- and
Bino-dressed operators vanish by Fierz identity under the universality
assumption \cite{matt}.

Charged Wino and both charged and neutral Higgsinos contribute to
significantly more channels due to the possibility for
generation-mixing through unitary matrix and Yukawa factors,
respectively, although they come at the price of suppression from
off-diagonal elements of pairs of those factors, {\it e.g.}
$U^d_{k2}\, U^u_{l1}$ or $y^u_{\, l1}\, y^d_{\, l1}$. However, the
sparse or hierarchical texture of the GUT Yukawas leads to a very
large spread in the magnitudes of the $C_{ijkl}$ values, and so there
is no general dominance of the operators without such suppression
factors over those with them.

Higgsino-dressed operators will generally contribute through both
left- and right-handed channels, and interference is present among any
group with all external legs of matching chiralities. Because we are
deliberately choosing large $\tan \beta$ to examine the most general
scenario, there is no overall suppression of right-handed operators
present in the analysis. Hence, we will make no {\it a priori}
assumptions about the significance of their contributions, and have
rather examined all contributions explicitly.

Two remaining restrictions that apply here are: $(a)$ since the mass
terms for the $W$ and the SUSY Higgs couple $W^+$ to $W^-$ and $H_u$ to
$H_d$, one will not see contributions from triplet operators with
sfermions of like $SU(2)$ flavor, and $(b)$ one will not see the
triplet operator $\tilde{u}du\tilde{e}$ dressed by $\tilde{h}^\pm$ nor
$u\tilde{d}d\tilde{\nu}$ dressed by $\tilde{h}^0$ because each would
result in an outgoing left-handed anti-neutrino.

We can write the operator for any pertinent decay mode as a generic
Wino- or Higgsino coefficient times one of several flavor-specific
sub-operators; the forms of the general operators are

\begin{equation}
  \mathscr{O}_{\widetilde{W}} = \left(\frac{i \alpha_2}{4\pi}\right)
  \left(\frac{1}{M_\mathcal{T}}\right)\, I\left(M_{\widetilde{W}},
  m_{\tilde{q}}\right) \mathscr{C}_{\widetilde{W}}^{\mathcal{A}}
  \label{eq:Wop}
\end{equation}
and
\begin{equation}
  \mathscr{O}_{\tilde{h}} = \left(\frac{i}{16\pi^2}\right)
  \left(\frac{1}{M_\mathcal{T}}\right)\, I\left(\mu,
  m_{\tilde{q}}\right) \mathscr{C}_{\tilde{h}}^{\mathcal{A}},
  \label{eq:hop} \vspace{4mm}
\end{equation}
where\footnote{One might notice that this expression for $I(a,b)$
differs from what is usually given in the literature for analogous
proton decay expressions; the discrepancy is due to inclusion of
the universal mass assumption prior to evaluating the loop integral.
\vspace{2mm}}
\begin{equation}
  I(a,b) = \frac{a}{b^2\!-\!a^2} \left\{\,1\, +\,
  \frac{a^2}{b^2\!-\!a^2} \log\left(\frac{a}{b}\right) \right\},
  \nonumber \vspace{4mm}
\end{equation}
and the sub-operators $\mathscr{C}^{\mathcal{A}}$ are, for the neutral
Wino:\,\footnote{We include the spinor and color details for the
neutral Wino operators, but suppress them afterward.}
\begin{align}
  \mathscr{C}_{\widetilde{W}^0}^I &= ~\epsilon_{abc}
  (u^{T\,a}\, C^{-1}\, s^c)\, \widehat{C}^L_{1[12]l}\, 
  (d^{T\,b}\, C^{-1}\, \nu_l) \nonumber \\
  \mathscr{C}_{\widetilde{W}^0}^{I\!I} &= ~\epsilon_{abc}
  (u^{T\,a}\, C^{-1}\, d^b)\, \widehat{C}^L_{1[12]l}\, 
  (s^{T\,c}\, C^{-1}\, \nu_l)
  \label{eq:CW0ops} 
\end{align}

%\newpage
\noindent for the (charged) Wino:
\begin{align}
  \mathscr{C}_{\widetilde{W}^\pm}^I &= \frac{1}{2} (u\, d_j)\,
  \widehat{C}^L_{[ij1]l}\, U^d_{ii'}\, U^\nu_{ll'}\, (d_{i'}\, \nu_{l'}) 
  \nonumber \\
  \mathscr{C}_{\widetilde{W}^\pm}^{I\!I} &= \frac{1}{2} (u\, e_l)\, 
  \widehat{C}^L_{[1jk]l}\, U^d_{kk'}\, U^u_{j1}\, (d_{k'}\, u) \nonumber \\
  \mathscr{C}_{\widetilde{W}^\pm}^{I\!I\!I} &= -\frac{1}{2} (u\, d_{k})\, 
  \widehat{C}^L_{1[jk]l}\, U^u_{j1}\, U^e_{ll'}\, (u\, e_{l'}) \nonumber \\
  \mathscr{C}_{\widetilde{W}^\pm}^{I\!V} &= -\frac{1}{2} (d_j\, \nu_l)\, 
  \widehat{C}^L_{i[jk]l}\, U^d_{ii'}\, U^u_{k1}\, (d_{i'}\, u)
  \label{eq:CWpops}
\end{align}
for the charged Higgsino: 
\begin{align}
  \mathscr{C}_{\tilde{h}^\pm}^{I} &= (u\, e_l)\, 
  \widehat{C}^L_{[1jk]l}\: y^{d\, \dagger}_{kk'}\, y^{u\, \dagger}_{j1}\, 
  (d^{\,\mathcal{C}}_{k'}\, u^{\mathcal{C}}) \nonumber \\ 
  \mathscr{C}_{\tilde{h}^\pm}^{I\!I} &= -(u\, d_{k})\,
  \widehat{C}^L_{1[jk]l}\: y^{u\, \dagger}_{j1}\, y^{e\, \dagger}_{ll'}\, 
  (u^{\mathcal{C}}\, e^{\mathcal{C}}_{l'}) \nonumber \\
  \mathscr{C}_{\tilde{h}^\pm}^{I\!I\!I} &= -(d_j\, \nu_l)\, 
  \widehat{C}^L_{i[jk]l}\: y^{d\, \dagger}_{ii'}\, y^{u\, \dagger}_{k1}\,
  (d^{\,\mathcal{C}}_{i'}\, u^{\mathcal{C}}) \nonumber \\ 
  \mathscr{C}_{\tilde{h}^\pm}^{I\!V} &= (u^{\mathcal{C}}\, 
  d^{\,\mathcal{C}}_j)\, \widehat{C}^R_{[ij1]l}\: y^u_{ii'}\, y^e_{ll'}\, 
  (d_{i'}\, \nu_{l'}) \nonumber \\
  \mathscr{C}_{\tilde{h}^\pm}^V &= (u^{\mathcal{C}}\, 
  e^{\mathcal{C}}_l)\, \widehat{C}^R_{[1jk]l}\: y^u_{kk'}\, y^d_{j1}\,
  (d_{k'}\, u)
  \label{eq:Chpops}
\end{align}
and for the neutral Higgsino:
\begin{align}
  \mathscr{C}_{\tilde{h}^0}^{I} &= -(u\, d_{k})\,
  \widehat{C}^L_{[ij1]l}\: y^{u\, \dagger}_{i1}\, y^{e\, \dagger}_{ll'}\, 
  (u^{\mathcal{C}}\, e^{\mathcal{C}}_{l'}) \nonumber \\
  \mathscr{C}_{\tilde{h}^0}^{I\!I} &= -(u\, e_l)\,
  \widehat{C}^L_{[1jk]l}\: y^{d\, \dagger}_{kk'}\, y^{u\, \dagger}_{j1}\, 
  (d^{\,\mathcal{C}}_{k'}\, u^{\mathcal{C}}) \nonumber \\ 
  \mathscr{C}_{\tilde{h}^0}^{I\!I\!I} &= (d_j\, \nu_l)\, 
  \widehat{C}^L_{i[jk]l}\: y^{u\, \dagger}_{i1}\, y^{d\, \dagger}_{kk'}\, 
  (u^{\mathcal{C}}\, d^{\,\mathcal{C}}_{k'}) \nonumber \\ 
  \mathscr{C}_{\tilde{h}^0}^{I\!V} &= -(u^{\mathcal{C}}\, 
  d^{\,\mathcal{C}}_j)\, \widehat{C}^R_{[ij1]l}\: y^u_{i1}\, y^e_{ll'}\,
  (u\, e_{l}) \nonumber \\
  \mathscr{C}_{\tilde{h}^0}^V &= -(u^{\mathcal{C}}\, 
  e^{\mathcal{C}}_l)\, \widehat{C}^R_{[1jk]l}\: y^u_{k1}\, y^d_{jj'}\,
  (u\, d_{j'}) 
  \label{eq:Ch0ops}
\end{align} %\newpage
\noindent Again the hats on $\widehat{C}^{L,R}$ indicate
$\hat{h},\hat{f},\hat{g}$ are rotated to the mass basis, which I will
discuss in detail shortly. Note that $UDUE$ and $UDD\mathcal{N}$
operators generally differ by a sign, as do diagrams dressed by
$\tilde{h}^\pm_{u,d}$ and $\tilde{h}^0_{u,d}$. These sign differences
lead to cancellations within the absolute squared sums of interfering
diagrams, and even cancellation of entire diagrams with each other in
some cases.

The corresponding Feynman diagrams for all non-vanishing channels 
for the $K^+ \bar{\nu}_l$, $K^0 \ell^+$, $\pi^+ \bar{\nu}_l$, and
$\pi^0 \ell^+$ modes are catalogued in the Appendix of
\cite{matt}. All non-negligible contributions used here in the
analysis are catalogued in Appendix B.

Note that since the SUSY Yukawas $y^f$ present in the
$\mathscr{C}^{\mathcal{A}}$ are not physically determined, we define
approximations to the Yukawas by using weak scale masses rotated by
GUT-scale $U^f$ (which are determined by our fermion sector fitting):
\begin{equation}
  y^{u} = \frac{1}{v_u}\: U_u\, \left(\mathcal{M}^{\rm wk}_u\right)^D\,
  U_u^\dagger, \nonumber
\end{equation}
where $v_u = v_{\rm wk} \sin \beta$, or, in component notation,
\begin{equation}
  y^{u}_{ij} = \frac{1}{v_u}\: \sum\limits_k\, m^u_k\, U^u_{ik}\,
  U^{u\,*}_{jk},
\end{equation}
with similar expressions for $y^{d}$ and $y^{e}$.

We used mass values from the current PDG \cite{pdg}; light masses were
run to the 1-GeV scale, and top and bottom masses were taken on-shell.

Note that since the Yukuwa factors always appear in pairs of opposite
flavor in the Higgsino operators, and since
$\frac{1}{\sin\beta\cos\beta} \simeq \tan\beta$ for large $\beta$, the
Higgsino contributions to proton decay $\sim \frac{\tan^2\beta}{v_{\rm
wk}^4}$ for this model.

\subsection{Rotation to Mass Basis} There are generally two
distinct mass-basis rotations possible for each of the
$UDUE\,\mbox{-}$, $UDD\mathcal{N}$-, and $U^{\mathcal{C}}
D^{\mathcal{C}} U^{\mathcal{C}} E^{\mathcal{C}}$-type triplet
operators; the difference between the two depends on whether the
operator is ``oriented'' ({\it i.e.}, in the diagram) such that the
lepton is a scalar. For a given orientation, a unitary matrix
corresponding to the fermionic field at one vertex in the triplet
operator will rotate every coupling present in ${C}^{L,R}$ pertaining
to that vertex; an analogous rotation will happen for the other vertex
in the operator. For example, looking at the $\pi^+ \bar{\nu_l}$
channel in Figure \ref{fig:compfd}(a), every coupling
$\lambda_{ij}~(\lambda = h,f,g)$ from $C^L_{ijkl}$ present at the
$\tilde{\phi}_{\mathcal{T}}$ vertex will be rotated by $U^d$;
similarly all $\kappa_{kl}$ present at the
$\tilde{\phi}_{\overline{\mathcal{T}}}$ vertex will be rotated by
$U^u$.

The down quark field shown is a mass eigenstate quark resulting from
unitary the rotation, which we can interpret as a linear combination
of flavor eigenstates: $d_j = U^d_{jm}\, d'_{m}$, with $j=1$; applying
the same thinking to the up quark, we can also write $u_k^T =
u'^T_{p}\, U^{u\,T}_{pk}$, with $k=1$. To work out the details of the
rotations, we can start with the $d=5$ operator written in terms of
flavor states\footnote{Recall the scalars are both mass and flavor
eigenstates under the universal mass assumption}, $\sum\nolimits_a x_a
(\tilde{u}_i\, \lambda^a_{im}\, d'_m) (u'_p \kappa^a_{pl}\,
\tilde{e}_l)$, where I have expanded $C^L_{impl}$ in terms of its
component couplings and chosen the indices with the malice of
forethought; now we can write
\begin{align}
  &\sum\limits_a x_a (\tilde{u}^T_i\, C^{-1} \lambda^a_{im}\, d'_{m})
  (u'^{\,T}_p\, \kappa^a_{pl}\, C^{-1}\, \tilde{e}_l) \nonumber
\end{align}
\begin{align}
  = ~ &\sum\limits_a x_a (\tilde{u}^T_i\, C^{-1}
  \underbrace{\lambda^a_{im}\, U^{d\, \dagger}_{mj}}_{\displaystyle
  \equiv \hat{\lambda}^a_{ij}}\, \underbrace{U^d_{jn}\,
  d'_{n}}_{\displaystyle d_j}) (\underbrace{u'^{\,T}_p\,
  U^{u\,T}_{pk}}_{\displaystyle u_k^T}\, \underbrace{U^{u\,*}_{kq}\,
  \kappa^a_{ql}}_{\displaystyle \equiv \hat{\kappa}^a_{kl}}\,
  C^{-1}\, \tilde{e}_l). \nonumber
\end{align}
Using the new definitions for $\hat{\lambda}$, we can see that the
rotated coefficient $\widehat{C}^L$ corresponding to the expression in
eq.\,(\ref{eq:Cs}) has become
\begin{align}
  \widehat{C}^L_{ijkl} &= x_0 \hat{h}_{ij} \hat{h}_{kl} + x_1
  \hat{f}_{ij} \hat{f}_{kl} - x_3 \hat{h}_{ij} \hat{f}_{kl} + \dots
  \nonumber \\ 
  &= x_0 (h\, U_d^\dagger)_{ij} (U_u^* h)_{kl}\, +\, x_1
  (f\, U_d^\dagger)_{ij} (U_u^* f)_{kl}\, -\, x_3 (h\,
  U_d^\dagger)_{ij} (U_u^* f)_{kl}\, +\, \dots
\end{align}
Note that this version of $\widehat{C}^L$ is only valid for
$UDUE$-type operators with this orientation in the diagram, namely,
those with a scalar $\tilde{e}$; there is an analogous pair of
rotations for $UDUE$ with a scalar down and fermionic lepton, as well
as two each for $UDD\mathcal{N}$ and $U^{\mathcal{C}} D^{\mathcal{C}}
U^{\mathcal{C}} E^{\mathcal{C}}$, for a total of six possible schemes.

\subsection{Proton Decay Width}
The full proton decay width $\tau \left(\, p \rightarrow \mathrm{M}
\bar\ell \, \right)$ (where M\:$= K, \pi$ is the final meson state) 
can be written as
\begin{equation}
  \Gamma = \frac{1}{4\pi}\, \beta_H^2\, (A_L\,A_S)^2 \left( \lvert
  \mathscr{O}_{\widetilde{W}} \rvert^{\,2} + \lvert
  \mathscr{O}_{\tilde{h}} \rvert^{\,2} \right)\; \mathrm{p},
  \label{eq:gtotOs}
\end{equation}
where
\begin{itemize}
  \item $\mathscr{O}_{\widetilde{W}}$ and $\mathscr{O}_{\tilde{h}}$
    are the operators given in (\ref{eq:Wop}) and (\ref{eq:hop}).
  \item The Hadronic factor $\beta_H^2$ is defined by $\bra{\mathrm{M}} 
    (qq)q \ket{p} \sim \beta_H P u_p$; the parameter is discussed
    extensively in works such as \cite{fukugita} and \cite{claudson};
    its value is now most commonly found in the range (0.006 - 0.03),
    with a tendency to prefer $\beta_H \sim 0.015$, as seen in
    \cite{fukugita}. We will take a slightly favorable approach and
    use $\beta_H = 0.008$.
  \item The factors $A_L$ and $A_S$ account for the renormalization
    of the $d=6$ dressed operators from $M_p \rightarrow M_{SUSY}$ and
    $M_{SUSY} \rightarrow M_U$, respectively; their values have been
    calculated in the literature as $A_L = 0.4$ and $A_S = 0.9
    \mbox{-} 1.0$ \cite{hisano}.
  \item In the rest frame of the proton, the external momentum p
    $= \lvert \mathbf{p} \rvert \equiv -\mathbf{p}_{\mathrm{M}} 
    = \mathbf{p}_\ell$ is given by
    \begin{equation}
      \mathrm{p} \, \simeq \, \frac{M_p}{2}
      \left(1 - \frac{m^2_{\mathrm{M}}}{M_p^2} \right),
    \end{equation}
    where we have assumed $m_\ell^2 \ll \lvert \mathbf{p}
    \rvert^{\,2}$ (which is only marginally valid for $m_\mu$ but
    clearly so for $m_e$). Note that p $\sim M_p/2$ for pion modes,
    but that value is reduced by a factor of $\sim$\,25\% for kaon
    modes.
\end{itemize}

Let us make one further definition to allow for clear statement of the 
working formulae for the partial decay widths of the proton. We define
$\mathrm{C}^{\mathcal{A}}$ as extended forms of the $C_{ijkl}$
by
\begin{align}
  \mathscr{C}_{\widetilde{W}}^{\mathcal{A}} =
  \mathrm{C}_{\widetilde{W}}^{\mathcal{A}} (qq)(q\ell) \nonumber \\
  \mathscr{C}_{\tilde{h}}^{\mathcal{A}} =
  \mathrm{C}_{\tilde{h}}^{\mathcal{A}} (qq)(q\ell) 
\end{align}
so that these coefficients contain the $U^f$ or $y^f$ factors as well
as the $C_{ijkl}$ of the $\mathscr{C}^{\mathcal{A}}$ operators in
(\ref{eq:CW0ops})-(\ref{eq:Ch0ops}). This gives us the ability to easily
translate an operator expression like \footnote{Here, we use the
approximation $I\left(M_{\widetilde{W}}, m_{\tilde{q}}\right) \sim
M_{\widetilde{W}} / m_{\tilde{q}}^2$ for the loop integral factor.}
\begin{equation}
  \mathscr{O}_{\widetilde{W}} (K^+ \bar{\nu}) \simeq 
  \left(\frac{i \alpha_2}{4\pi}\right) \frac{1}{M_\mathcal{T}}\,
  \left(\frac{M_{\widetilde{W}}}{m_{\tilde{q}}^2}\right)
  \{\mathscr{C}_{\widetilde{W}}^I + \mathscr{C}_{\widetilde{W}}^{I\!V}
  \}
  \label{eq:OKnuW}
\end{equation}
into a partial decay width statement,
\begin{equation}
  \Gamma_{\widetilde{W}} (p \rightarrow K^+ \bar{\nu}) \simeq
  \frac{1}{4\pi} \left(\frac{\alpha_2}{4\pi}\right)^2
  \frac{1}{M^2_{\mathcal{T}}} \left(\frac{M_{\widetilde{W}}}
  {m^2_{\tilde{q}}}\right)^2 \beta_H^2\,(A_L A_S)^2\, \mathrm{p}\:
  \lvert\, \mathrm{C}_{\widetilde{W}}^I +
  \mathrm{C}_{\widetilde{W}}^{I\!V}\, \rvert^{\,2}, 
  \label{eq:gKnuW}
\end{equation}
without losing either information or readability. Hence, we can now
present relatively compact and intelligible expressions for the Wino-
and Higgsino-dressed partial decay widths of the proton for generic
mode $p \rightarrow \mathrm{M} \bar\ell$:
\begin{align}
  \label{eq:gammaW}
  \Gamma_{\widetilde{W}} (p \rightarrow \mathrm{M} \bar\ell) \simeq
  \frac{1}{4\pi} \left(\frac{\alpha_2}{4\pi}\right)^2
  \frac{1}{M^2_{\mathcal{T}}} \left(\frac{M_{\widetilde{W}}}
  {m^2_{\tilde{q}}}\right)^2 \beta_H^2\,(A_L A_S)^2\, \mathrm{p}\:
  \Big\lvert\! \sum\limits_{\mathcal{A} \in \mathrm{M} \bar\ell}\!
  \mathrm{C}_{\widetilde{W}}^{\mathcal{A}}\, \Big\rvert^{\,2} \\
  \Gamma_{\tilde{h}} (p \rightarrow \mathrm{M} \bar\ell) \simeq
  \frac{1}{4\pi} \left(\frac{1}{16\pi^2}\right)^2
  \frac{1}{M^2_{\mathcal{T}}} \left(\frac{\mu}
  {m^2_{\tilde{q}}}\right)^2 \beta_H^2\,(A_L A_S)^2\, \mathrm{p}\:
  \Big\lvert\! \sum\limits_{\mathcal{A} \in \mathrm{M} \bar\ell}\!
  \mathrm{C}_{\tilde{h}}^{\mathcal{A}} \Big\rvert^{\,2}. 
  \label{eq:gammah}
\end{align}
For the numerical analysis, we used the generic values 
\begin{equation*}
  M_{\cal T} = 2 \!\times\! 10^{16}\,{\rm GeV}, \quad M_{\widetilde{W}} 
  = \mu = 300\,{\rm GeV}, \quad m_{\tilde{q}} = 5\,{\rm TeV}. 
\end{equation*}
Like our choice for $\tan \beta$, these choices for $\mu$ and
$m_{\tilde{q}}$ are chosen to yield deliberately strict constraints in
order to hold the model to the highest feasible scrutiny; {\it i.e.},
we are presenting the ``worst case scenario'' for the model.

Also, let us repeat here that because of the two SUSY Yukawa coupling
factors in the $\mathrm{C}_{\tilde{h}}^{\mathcal{A}}$, which always
come in opposite flavor,
\begin{equation}
  \Gamma_{\tilde{h}} \propto \left(\frac{1}{v_{\rm wk}^2
  \sin\beta\cos\beta} \right)^2 \sim
  \frac{\tan^2\beta}{v_{\rm wk}^4}. \nonumber
\end{equation}

\vspace{2mm}
Finally, note that because the Higgsino vertices change the
chiralities of the outgoing fermions, there can be no interference
between Wino- and Higgsino-dressed diagrams, as suggested by
eq.\,(\ref{eq:gtotOs}); however, since diagrams for the right-handed
$C^R$ operators have outgoing {\it left-handed} fermions by the same
Higgsino mechanism, diagrams for $C^R$- and $C^L$-type operators with
the same external particles of matching chiralities {\it do}
interfere with each other, and so all such contributions to a given
mode do in fact go into the same absolute-squared sum factor, as
suggested by eq.\,(\ref{eq:gammah}). 
%\newpage

\section{Fermion Fitting and Leptonic CP Violation} \label{fit} 
Our first computational goal with this model was to understand its
preferences and flexibility in predicting a value for the neutrino
sector CP phase for experimental agreement with all known fermion
parameters. Because of the beginning trend in the data, we
placed some focus on phase values in the large negative range, {\it
i.e.} with $\sin \delta_{\rm CP} \sim 1$; however, given the current
uncertainty in the measured value, we also wanted to consider the
entire range of possibilities. 

By diagonalizing the mass matrices given in eq.\,(\ref{eq:mass}), with
the Yukawa textures shown in (\ref{eq:y}), one can obtain the
GUT-scale fermion masses and mixing angles for a given set of values
for the mass matrix parameters $h_{ij}$, $f_{ij}$, $r_i$, etc. To find
a best fit to the experimental data, we use the {\tt Minuit} tool
library for Python \cite{minuit,python} to minimize the sum of
chi-squares for neutrino mass-squared differences $\Delta m_{21}^2$
(aka $\Delta m_\odot^2$) and $\Delta m_{32}^2$ (aka $\Delta m_{\rm
atm}^2$) and the PMNS mixing angles as well as the mass eigenvalues
and CKM mixing angles in the charged-fermion sector.

\subsubsection{Numerical Methods.}\label{numerical} Because the input
parameter space of the model is large and highly non-linear, and
because our minimization tools depend on a measurement of the local
gradient, our choices for initial values in any search are likely to
affect the output results.  In order to ensure a high likelihood for
finding global minima, after setting some rough initial values for the
inputs using analytical arguments, we use an iterative systematic
approach in refining those values.

Furthermore, for this analysis, since we were interested in the full range of
possibilities for the CP phase output, we chose multiple initial
positions within the parameter space from which to perform our
minimization. Our specific choices result in three distinct values for
the CP phase while approximating the remaining fermion sector to a
crude but sufficient degree for initialization.

The three initial inputs are given in Table \ref{table:initsI}.

From each starting position, we search for a series of potential fits by
iteratively including a ``target'' value for the PMNS Jarlskog
invariant from the interval \\ $J_\nu = [- 0.03466, 0.03466]$, which
corresponds to the entire range of possible phase values $-1 \leq \sin
\delta_{CP} \leq 1$. The target is implemented by including $J_\nu$ in
the sum of chi-squares for the outputs. Once the best fit is found,
this hypothetical $J_\nu$ contribution to the sum of chi-squares is
subtracted out to obtain the true value for the fit.

We fit to Type-I and type-II seesaw neutrino masses separately and
so report the results for each accordingly. Note that throughout the
analysis, we take $v_u = 117.8$\,GeV, calculated with $\tan\beta = 55$
and with $v_{\rm wk}$ run to the GUT scale \cite{das}.
%The corresponding value for the down-type vev is $v_d = 2.26$\,GeV.

\begin{table}[t]
\begin{center}
  \begin{tabular}{||c|c|c|c||}\hline\hline
    & \blue{Init \#1} & \green{Init \#2} & \red{Init \#3} \\\hline
    $M$ (GeV)            & 80.2  & 76.1   & 79.0   \\ 
    $\tilde f_{11}$ (GeV)& 0.0055& 0.01013& 0.0145 \\  
    $\tilde f_{12}$ (GeV)& 0.0965& -0.089 & 0.064  \\ 
    $\tilde f_{13}$ (GeV)& 0.608 & 0.9397 & 1.55   \\  
    $\tilde f_{22}$ (GeV)& 1.094 & 0.866  & 1.32   \\  
    $\tilde f_{23}$ (GeV)& 1.21  & 1.4884 & -0.75  \\  
    $\tilde f_{33}$ (GeV)& 1.51  & 3.55   & -3.95  \\  
    $\tilde g_{12}$ (GeV)& 0.26  & 0.20   & 0.359  \\ 
    $\tilde g_{13}$ (GeV)& 0.08  & 0.0535 & 0.013  \\  
    $\tilde g_{23}$ (GeV)& 0.178 & 0.35   & -0.01  \\  
    $r_1/ \tan\beta$     & 0.0215& 0.0247 & 0.0175 \\  
    $r_2$                & 0.191 & 0.24414& 0.159  \\  
    $r_3$                & 0.0108& 0.006  & 0.0213 \\  
    $c_e$                & 0.355 & -3.328 & -3.05  \\  
    $c_\nu$              & 153.87& 45.218 & 127.0  \\  
    $\delta m_b$ (GeV)   & -24.5 & -28.0  & -10.25 \\   
    $\delta V_{cb}$ (GeV)&   0.88& 0.515  & 1.07   \\   
    $\delta V_{ub}$ (GeV)& -0.195& -0.844 & -1.0   \\   
    \hline\hline                                 
  \end{tabular}
  \caption{\footnotesize Initial values for the model input parameters
  at the GUT scale with type-I seesaw. Label colors correspond to
  those in the plots to follow.}
\label{table:initsI}
\end{center}
\end{table}

\subsubsection{A Note on Threshold Corrections.} Threshold corrections
at the SUSY scale are $\propto\tan\beta$, and so should be large in
this analysis \cite{poko}. The most substantial correction is to the
bottom quark mass, which is dominated by gluino and chargino loop
contributions; this correction also induces changes to the CKM matrix
elements involving the third generation. The explicit forms of these
corrections can be seen in a previous work on a related model
\cite{matt2}. Additionally, smaller off-diagonal threshold corrections
to the third generation parts of $\mathcal{M}_d$ result in small
corrections to the down and strange masses as well as further
adjustments to the CKM elements. All such corrections can be
parametrized in the model by
\begin{align}
  \mathcal{M}'_d = \mathcal{M}_d + 
    \frac{r_1}{\tan\beta}\left(\begin{array}{ccc}
    0 & 0 & \delta V_{ub}\\ 0 & 0 & \delta V_{cb}\\
    \delta V_{ub} & \delta V_{cb} & \delta m_b \end{array}\right),
\end{align}
\smallskip where $\mathcal{M}_d$ is given by eq.\,(\ref{eq:mass}). If
we simply take this augmented form for $\mathcal{M}_d$ as part of the
model input, the $\delta$ parameters can be taken as free parameters,
up to some constraints, which can then aid in the fitting. Their fit
values induce constraints on the Higgs and the light stop and sbottom
masses. The implications of this implementation were considered in
detail for a related model in \cite{matt2}; in comparing to that work,
one can determine for this model that large $\tan\beta$ and relatively
small threshold corrections result in weak and less interesting
constraints on the Higgs and squark masses, so we will not consider
them further in this analysis.

%it is of the form \begin{equation} \frac{\delta m_b}{m_b} \simeq
%\epsilon_1 + \epsilon_2|V_{tb}|^2, \end{equation} where
%\begin{align*} \epsilon_1 = \frac{2\alpha_s}{3\pi}\,\mu
%m_{\tilde{g}}\tan\beta~ I_3 (m^2_{\tilde{g}},m_{{\tilde b}_1}^2,m_{
%{\tilde b}_2}^2) \label{eq:eps1}\\ \epsilon_2 =
%\frac{1}{16\pi^2}\,\mu A_t y_t^2 \tan\beta~ I_3(\mu^2, m^2_{ {\tilde
%t}_1}, m^2_{ {\tilde t}_2}) \label{eq:eps2} \end{align*} and the loop
%integral function $I_3$ is \begin{equation} I_3(a,b,c) =
%\frac{ab\log\left(\frac{a}{b}\right) + bc
%\log\left(\frac{b}{c}\right) + ca \log\left(\frac{c}{a}\right)}
%{(a-b)(b-c)(a-c)}. \nonumber \end{equation} This correction induces
%changes to CKM matrix elements involving the third generation of the
%form \begin{equation} \frac{\delta V_{ub}}{V_{ub}} \simeq
%\frac{\delta V_{cb}}{V_{cb}} \simeq \frac{\delta V_{td}}{V_{td}}
%\simeq \frac{\delta V_{ts}}{V_{ts}} \simeq -\epsilon_2.
%\end{equation}

\subsection{Fit Results for Type I Seesaw} If one takes $v_R
\lsim 10^{16}$\,GeV and $v_L \ll 1$\,eV, then the type-I
contribution is dominant over the type-II contribution, and
eq.\,(\ref{eq:neu}) becomes
\begin{equation}
  {\cal M}_\nu \simeq - {\cal M}_{\nu_D}\left(v_R f\right)^{-1}\left(
  {\cal M}_{\nu_D}\right)^T,
\end{equation}
In our initial searches, several similar fermion sector fits were
obtained with Dirac phase values of $\delta_{CP} \sim -65^\circ$.
Table \ref{table:paramsI} gives the values for the adjusted model
input parameters for one example of such a type-I fit, and Table
\ref{table:fitI} gives the corresponding output values for the fermion
parameters.
%Fits in this swath of parameter space still have $c_\nu \sim 50$ and
%$\delta m_b \sim 25$\,GeV, but this value for $c_\nu$, while slightly
%strange, is quite readily accommodated by the doublet mixing
%parameters, and such a value for the largest SUSY threshold
%correction is actually quite moderate for large $\tan\beta$.
The precise value for the $\overline{\bf 126}$ vev used in this fit is
$v_R = 1.21\! \times\! 10^{15}$\,GeV.

%Note also that the top and strange masses are quite a bit lower than
%in the type-II fit; however, note I have also quoted different
%experimental values with which agreement is maintained. The
%differences here come from an update to the work in \cite{das} in
%determining two-loop MSSM RGEs for fermion masses. The update
%\cite{bora} reports notably lower masses for all the quarks at
%$\tan\beta = 55$ and $\mu = 2.0 \!\times\! 10^{16}$\,GeV, especially
%for the up, down, strange, and top masses, due to updates in initial
%values and methodology. Hence, one should not give the specific values
%too much weight in such a fit, and I do not consider the reported
%differences to be significant. This same thinking applies for the
%type-II down mass value in Table \ref{table:fitII}.
%\vspace{4mm}

\begin{table}[b]
\begin{center}
  \begin{tabular}{||c|c||c|c||}\hline\hline
    $M$ (GeV) &           79.33& $r_1/ \tan\beta$ &      0.023735\\
    $\tilde f_{11}$ (GeV) &   0.014373& $r_2$ &                  0.2022 \\
    $\tilde f_{12}$ (GeV) &   0.097515& $r_3$ &                  0.0199 \\
    $\tilde f_{13}$ (GeV) &    0.56460& $c_e$ &                  -1.4963\\
    $\tilde f_{22}$ (GeV) &     1.0219& $c_\nu$ &                 24.317\\
    $\tilde f_{23}$ (GeV) &     0.0926& $\delta m_b$ (GeV) &     -28.00 \\
    $\tilde f_{33}$ (GeV) &    3.85295& $\delta V_{cb}$ (GeV) & -1.495\\
    $\tilde g_{12}$ (GeV) &    0.26397& $\delta V_{ub}$ (GeV) &  0.21557\\
    $\tilde g_{13}$ (GeV) &    0.22252& & \\
    $\tilde g_{23}$ (GeV) &    1.02642& & \\
    \hline\hline
  \end{tabular}
  \caption{\footnotesize Best-fit values for the model parameters at the
  GUT scale for a sample model fit with type-I seesaw.}
\label{table:paramsI}
\end{center}
\end{table}

\begin{table}[t!]
\begin{center}
  \begin{tabular}{||c|c|c||c|c|c||}\hline\hline
    & best fit & exp value & & best fit & exp value\\ \hline
    $m_u$ (MeV) &       0.7246 &    $0.72^{+0.12}_{-0.15}$ &
    $V_{us}$ &          0.22427&    $0.2243\pm 0.0016$\\
    $m_c$ (MeV) &        208.6 &    $210.5^{+15.1}_{-21.2}$ &
    $V_{ub}$ &           0.0030&    $0.0032\pm 0.0005$\\
    $m_t$ (GeV) &        80.113&    $80.45^{+2.9\,*}_{-2.6}$ &
    $V_{cb}$ &          0.03497&    $0.0351\pm 0.0013$\\
    $m_d$ (MeV) &        1.515 &    $0.930\pm 0.38^*$ &
    $J \times 10^{-5}$ & 2.29  &    $2.2\pm 0.6$\\
    $m_s$ (MeV) &        24.47 &    $17.6^{+4.9\,*}_{-4.7}$ &
    $\Delta m_{21}^2 / \Delta m_{32}^2$ & 0.0308 & $0.0309\pm 0.0015$\\
    $m_b$ (GeV) &        1.311 &    $1.24\pm 0.06^*$ &
    $\theta_{13}~(^\circ)$ & 9.397 &$8.88\pm 0.385$\\
    $m_e$ (MeV) &       0.3565 &    $0.3565^{+0.0002}_{-0.001}$ &
    $\theta_{12}~(^\circ)$ & 33.62 &$33.5\pm 0.8$\\
    $m_\mu$ (MeV) &      75.297&    $75.29^{+0.05}_{-0.19}$ &
    $\theta_{23}~(^\circ)$ & 43.79 &$44.1\pm 3.06$\\
    $m_\tau$ (GeV) &     1.61  &    $1.63^{+0.04}_{-0.03}$ &
    $\boldsymbol{\delta_{\rm CP}}~(^\circ)$ & {\bf -67} & \\
    \hline
    & & & $\sum \chi^2$ & 3.16& \\
    \hline\hline
  \end{tabular}
  \caption{\footnotesize Best-fit values for the charged fermion
  masses, solar-to-atmospheric mass squared ratio, and CKM and PMNS
  mixing parameters for a sample fit with Type-I seesaw. The $1\sigma$
  experimental values are shown \cite{das} ($^*$\,-\,\cite{bora}),
  \cite{pdg}; masses and mixings are extrapolated to the GUT scale
  using the MSSM RGEs. Note that the fit values for the bottom quark
  mass and the CKM mixing parameters involving the third generation
  shown here include threshold corrections.} %\vspace{-8mm}
\label{table:fitI}
\end{center}
\end{table}

\begin{figure}[t]
  \begin{center}
    \textbf{$\boldsymbol{\sum\chi^2}$ vs $\boldsymbol{\delta_{\rm CP}}$ 
    for Type I Seesaw}
    \includegraphics[width=10cm]{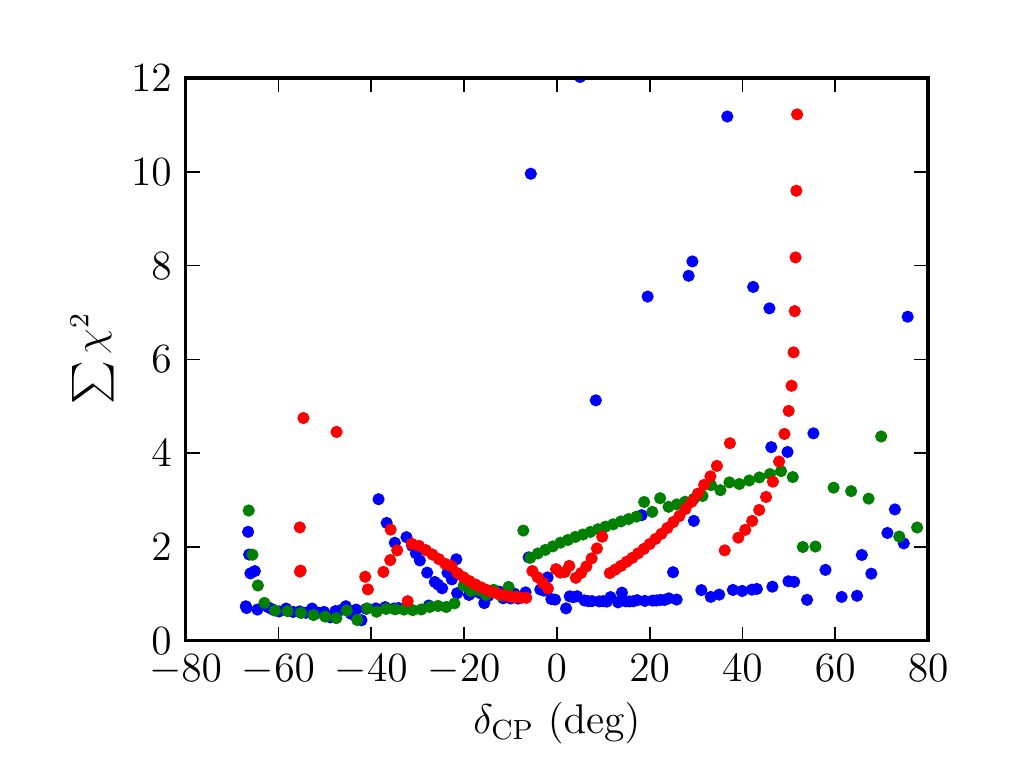}
    \caption{\footnotesize Comparisons of the Dirac CP phase angle to
    sum of chi squares for the model in the type-I seesaw case. Each
    color represents results for a different set of initial values for
    the input parameters.}
    \label{fig:X2CPI}
  \end{center}
\end{figure}

\begin{figure}[t]
  \begin{center}
    \textbf{Reactor and Atmospheric Angles vs $\boldsymbol{\delta_{\rm
    CP}}$ for Type I Seesaw} \\[1em]
    \includegraphics[width=7.6cm]{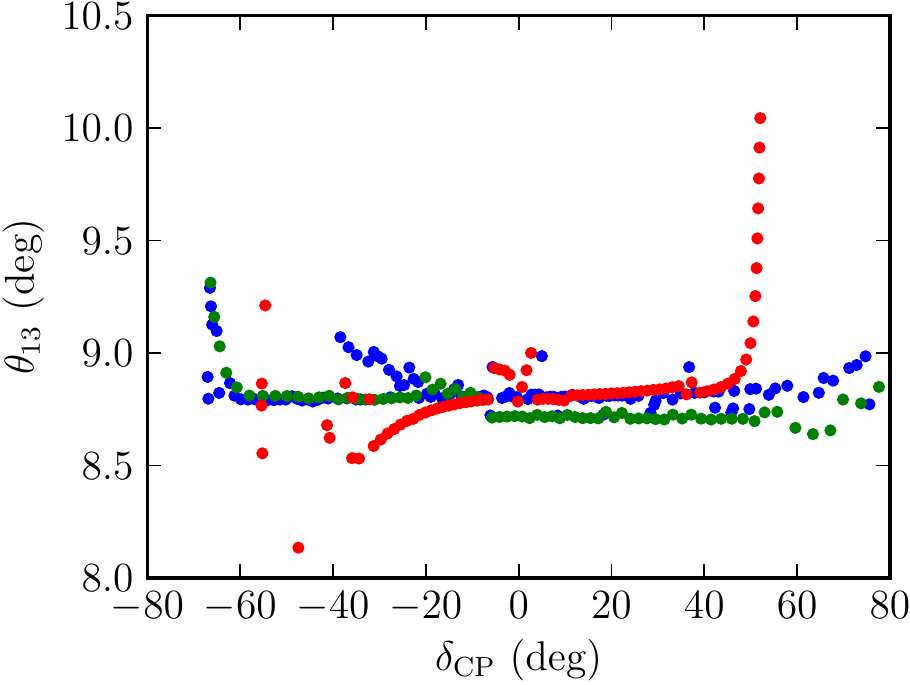}
    \includegraphics[width=7.35cm]{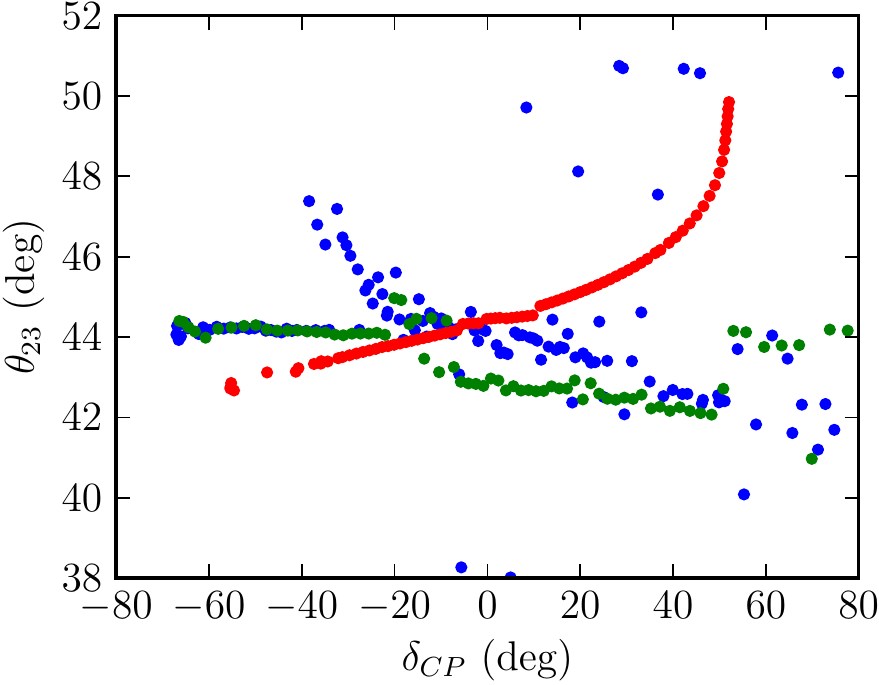}
    \caption{\footnotesize Comparisons of the Dirac CP phase angle to
    neutrino mixing angles $\theta_{13,23}$ the model in the type-I
    seesaw case. Each color represents results for a different set of
    initial values for the input parameters.}
    \label{fig:1323CPI}
  \end{center}
\end{figure}

To further explore the nature of CP phase in the model, we created
scatter plots of sum of chi-squares vs CP phase for the search methods
described above in \ref{numerical}. From the same data, we also made
plots of neutrino mixing angles $\theta_{13}$ and $\theta_{23}$ vs
corresponding phase value. These plots are given in Figures
\ref{fig:X2CPI} and \ref{fig:1323CPI}.

We can see from these results that in principle, the model can
accommodate virtually any value for the CP phase in the range
$-70^\circ \leq \delta \leq +80^\circ$ while still satisfying all
experimental constraints. Closer inspection shows however certain
preferred values or ranges, including large negative phases $-60^\circ
\leq \delta \leq -40^\circ$ as well as $\delta \sim -15^\circ$. The
former is certainly part of a robust region of parameter space, which
was explored thoroughly and includes the sample fit given in Tables
\ref{table:paramsI} and \ref{table:fitI}.

%\newpage
Notably absent is the maximal negative value of $\sin \delta_{CP} =
-1$, which may have experimental implications in the coming years.

\vspace{1em}

%#We generated some Gaussian spread in the
%#fit results above by varying the input parameters $r_2$, $c_\nu$,
%#$g_{12,13}$, and the vev $v_R$, and then applied constraints requiring
%#the resulting fermion sector fit outputs to be within both 1$\sigma$
%#and 3$\sigma$ of the central values.
%#
%#The resulting plots are shown in Figure \ref{fig:}. We include
%#the unfiltered data (in blue) to display the ``shape'' of the
%#parameter space near the fit, and then display the 1$\sigma$
%#(3$\sigma$) constrained data in teal (magenta). One can clearly see a
%#trend in the constrained data displaying a linear relationship between
%#the mixing angles and the CP phase.
%#
%#Note the discontinuity in the parameter space for the large CP phase
%#solution. We find that for large enough values of $c_\nu$ and
%#$g_{12}$, the phase ``jumps'' to a small value; furthermore, the good
%#fit results are found immediately adjacent to the discontinuity,
%#suggesting a curious nature to the model.

%\newpage
\subsubsection{Raw Couplings contributing to Proton Decay} In order to
calculate the $C_{ijkl}$ proton decay coefficients, we need the
``raw'' Yukawa couplings, $h,f,g$, which are obtained from the
dimensionful couplings, $\tilde{h},\tilde{f},\tilde{g}$, from the
relationships given in eq.\,(\ref{eq:tildes}), which are obtained
directly from the fit; There is some freedom in the values of the
elements of the matrices $\,{\cal U}, \,{\cal V}$ from the
viewpoint of this predominantly phenomenological analysis, but they
are constrained by both unitarity and the ratios $r_i$ and $c_\ell$,
which have been fixed by the fermion fit. Again, see \cite{DMM} for
details, or see \cite{matt2} for an example of such a calculation. The
resulting dimensionless couplings corresponding to this type-I fit are
are \smallskip
\begin{gather}
  h = \left( \begin{array}{ccc}
     0 & &\\ & 0 & \\ & & 1.6838
  \end{array}\right) \qquad
  f = \left( \begin{array}{ccc}
     0.00019072 &  0.00129401 &  0.00749214 \\
     0.00129401 &  0.01356053 &  0.00122855 \\
     0.00749214 &  0.00122855 &  0.05112795
  \end{array} \right) \nonumber \\[4mm]
  g = i \left(\begin{array}{ccc}
     0          & 0.02975586 & 0.02508341 \\
    -0.02975586 & 0          & 0.11570158 \\
    -0.02508341 &-0.11570158 & 0        \end{array} \right) 
\label{eq:hfgI}
\end{gather} %\newpage
\smallskip

\noindent Here, we see $f_{11} \sim 0$, but $f_{12}$,
$f_{13}$, and $g_{12}$ are large enough to seem unfavorable for proton 
decay. In the end, these factors are suppressed by other means, which
we will discuss in Section \ref{pfit}.

\subsection{Fit Results for Type II Seesaw} If one takes the
$\overline{\mathbf{126}}$ SM-singlet vev $v_R \gsim 10^{17}$\,GeV
({\it i.e.},\;the GUT scale), and the triplet vev $v_L \sim 1$\,eV,
then the type-II contribution ($v_L$ term) in eq.\,(\ref{eq:neu})
dominates over the type-I contribution ($v_R$ term) by an average of
two orders of magnitude in the neutrino mass matrix; therefore
eq.\,(\ref{eq:neu}) reduces to
\begin{equation}
  {\cal M}_\nu \simeq v_L f
\end{equation}

Initial searches here were based on the fit from \cite{matt}, which
gave a small negative CP phase value. Tables \ref{table:paramsII} and
\ref{table:fitII} give that fit as a sample here, which uses a precise
value for the $\overline\Delta_L$ vev of $v_L = 1.316$\,eV. 
\vspace{1em}
%corresponding output values for the fermion parameters, with
%experimentally measured values included for comparison. Note that the
%down quark mass is seemingly a bit low, which seems to be a general
%feature in this model, but I will discuss in the next section why this
%is not a problem. The , which I chose to fix the overall neutrino mass scale at
%$m_3 \sim 0.05$\,eV.

\begin{table}[h]
\begin{center}
  \begin{tabular}{||c|c||c|c||}\hline\hline
    $M$ (GeV) &               106.6& $r_1/ \tan\beta$ &    0.014601\\ 
    $f_{11}$ (GeV) &      -0.045564& $r_2$ &              0.0090315\\
    $f_{12}$ (GeV) &       0.048871& $r_3$ &                 1.154 \\ 
    $f_{13}$ (GeV) &       -0.59148& $c_e$ &                -2.5342\\ 
    $f_{22}$ (GeV) &       -2.06035& $c_\nu$ &                  n/a\\     
    $f_{23}$ (GeV) &        -1.4013& $\delta m_b$ (GeV) &   -22.740\\
    $f_{33}$ (GeV) &       -1.40644& $\delta V_{cb}$ (GeV) & 1.2237\\
    $g_{12}$ (GeV) &       0.018797& $\delta V_{ub}$ (GeV) & 4.2783\\
    $g_{13}$ (GeV) &       -0.92510& & \\
    $g_{23}$ (GeV) &        -3.8353& & \\
    \hline\hline
  \end{tabular}
  \caption{\footnotesize Best fit values for the input parameters at the
  GUT scale for a sample fit with type-II seesaw (taken from
  \cite{matt}).}
\label{table:paramsII}
\end{center}
\end{table}
\begin{table}[b]
\begin{center}
  \begin{tabular}{||c|c|c||c|c|c||}\hline\hline
    & best fit & exp value & & best fit & exp value\\ \hline
    $m_u$ (MeV) &       0.7172 &    $0.72^{+0.12}_{-0.15}$ &
    $V_{us}$ &          0.2245 &    $0.2243\pm 0.0016$\\
    $m_c$ (MeV) &       213.8 &    $210.5^{+15.1}_{-21.2}$ &
    $V_{ub}$ &          0.00326&    $0.0032\pm 0.0005$\\
    $m_t$ (GeV) &       106.8  &    $95^{+69}_{-21}$ &
    $V_{cb}$ &          0.0349 &    $0.0351\pm 0.0013$\\
    $m_d$ (MeV) &       0.8827 &    $1.5^{+0.4}_{-0.2}$ &
    $J \times 10^{-5}$ & 2.38 &    $2.2\pm 0.6$\\
    $m_s$ (MeV) &       34.04 &    $29.8^{+4.18}_{-4.5}$ &
    $\Delta m_{21}^2 / \Delta m_{32}^2$ & 0.03065& $0.0309\pm 0.0015$\\
    $m_b$ (GeV) &       1.209 &    $1.42^{+0.48}_{-0.19}$ &
    $\theta_{13}~(^\circ)$ & 9.057 &$8.88\pm 0.385$\\
    $m_e$ (MeV) &       0.3565 &    $0.3565^{+0.0002}_{-0.001}$ &
    $\theta_{12}~(^\circ)$ & 33.01 &$33.5\pm 0.8$\\
    $m_\mu$ (MeV) &     75.297 &    $75.29^{+0.05}_{-0.19}$ &
    $\theta_{23}~(^\circ)$ & 47.70 &$44.1\pm 3.06$\\
    $m_\tau$ (GeV) &    1.635 &    $1.63^{+0.04}_{-0.03}$ &
    $\delta_{\rm CP}~(^\circ)$ &   -7.5 & \\
    \hline
    & & & $\sum \chi^2$ & 6.0 & \\
    \hline\hline
  \end{tabular}
  \caption{\footnotesize Best fit values for the charged fermion
  masses, solar-to-atmospheric mass squared ratio, and CKM and PMNS
  mixing parameters for a sample fit with Type-II seesaw. The $1\sigma$
  experimental values are also shown for comparison \cite{das},
  \cite{pdg}, where masses and mixings are extrapolated to the GUT
  scale using the MSSM renormalization group equations (RGEs). Note
  that the fit values for the bottom quark mass and the CKM mixing
  parameters involving the third generation shown here include the
  SUSY-threshold corrections} 
  \label{table:fitII} %\vspace{-4mm}
\end{center}
\end{table}

Performance of the type II model beyond this point was largely
unsuccessful. We were unable to find additional initial values for the
input parameters which generated significantly different CP phase
values. So moving forward with only one initial position in the
parameter space, the scatter plots of chi-squares sum vs CP and
corresponding plots for the PMNS angles vs CP reveal a bleak
situation, which illuminates the difficulty we had in locating
different starting points.

The plots in question are given in Figures \ref{fig:X2CPII} and
\ref{fig:1323CPII} below. One quickly notices a much more limited
picture of potential phase values. It would seem that the type II
model can support small values only for $\delta_{CP}$, with best fits
(notably superior to the sample fit) giving $\delta \sim \pm
15^\circ$.

\begin{figure}[h]
  \begin{center}
    \textbf{$\boldsymbol{\sum\chi^2}$ vs $\boldsymbol{\delta_{\rm CP}}$ 
    for Type II Seesaw}
    \includegraphics[width=9cm]{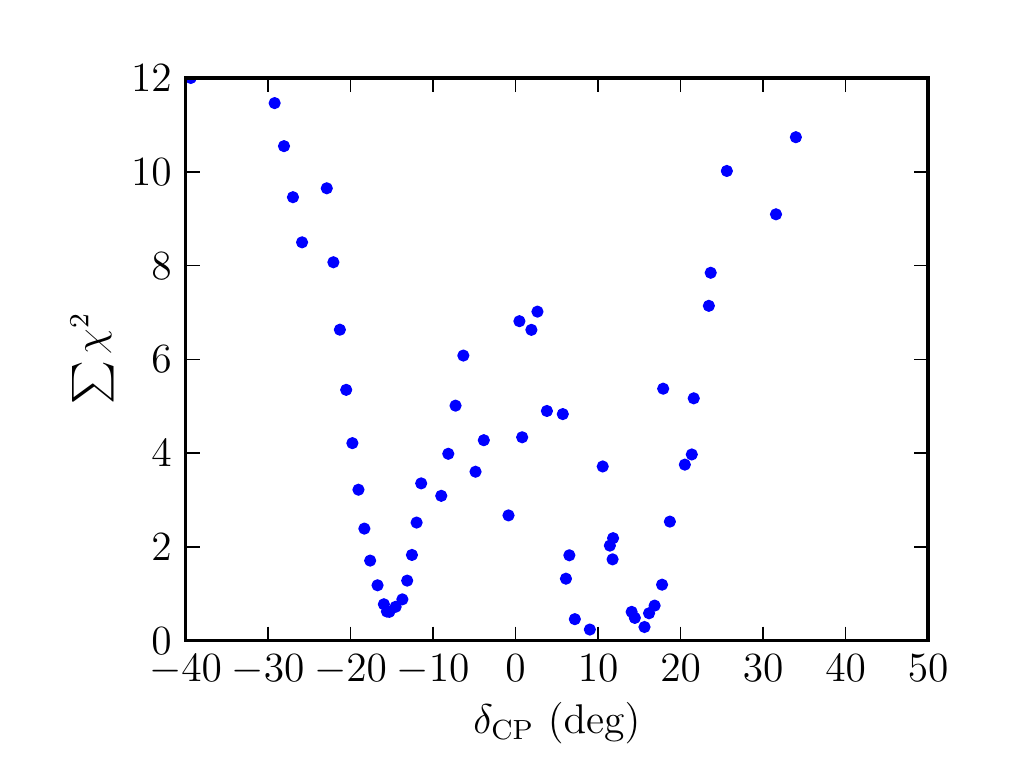}
    \caption{\footnotesize Comparisons of the Dirac CP phase angle to
    sum of chi squares for the model in the type-II seesaw case.}
    \label{fig:X2CPII} \vspace{-4mm}
  \end{center}
\end{figure}

\begin{figure}[b]
  \begin{center}
    \textbf{Reactor and Atmospheric Angles vs $\boldsymbol{\delta_{\rm
    CP}}$ for Type II Seesaw} \\[1em]
    \includegraphics[width=7.4cm,height=5.6cm]{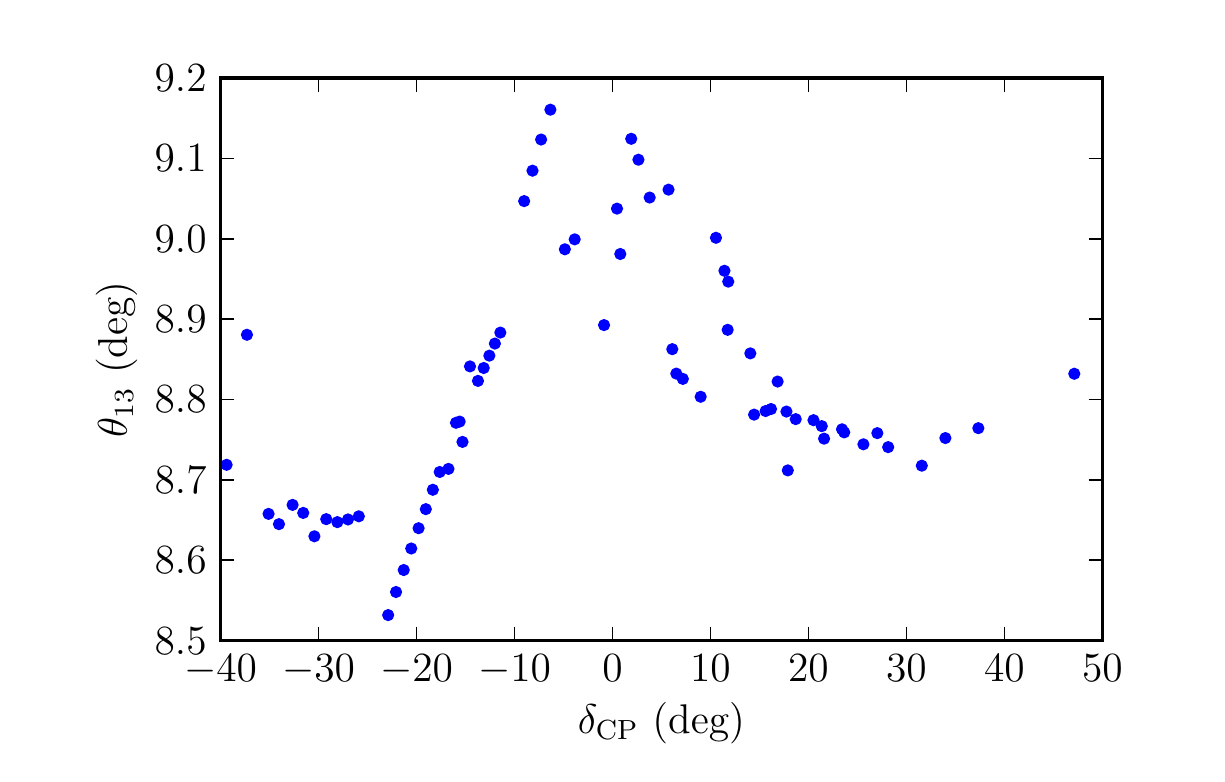}
    \includegraphics[width=7.4cm,height=5.6cm]{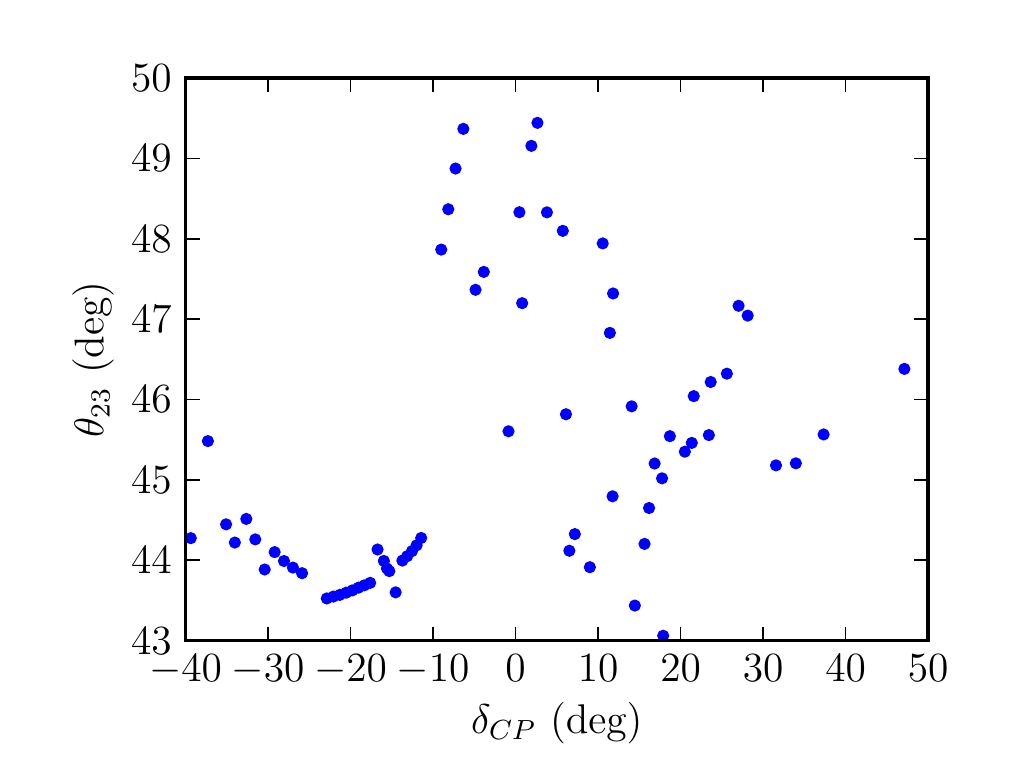}
    \caption{\footnotesize Comparisons of the Dirac CP phase angle to
    neutrino mixing angles $\theta_{13,23}$ for the model in the type-II
    seesaw case.}
    \label{fig:1323CPII}
  \end{center}
\end{figure}

As further evidence for the limitations here, note that the density of
the range of $J_\nu$ values explored on the interval [-0.03466, 0.03466]
was identical to that used in the red and green data for the type-I
case. The greatly reduced number of points seen in the plots is the
result of numerous minimization attempts failing to converge, as well
as additional convergences that gave $\sum \chi^2 \gsim 15$.

Given the highly limited results of this analysis, we can conclude with
statistical confidence that the {\it type-II-dominant version of this model
is unable to accommodate large CP phase values} of either sign, which
should result in a {\it definite fate} for the model as experiments close in
on a discovery.

\section{Proton Lifetime Correlation with Dirac CP Phase} \label{pfit}
In order to complete this analysis, we now turn to the relationship
between CP phase output and the lifetime of the proton. To obtain
values for the partial decay widths in consideration, we need values
for the $x_i$ and $y_i$ triplet mixing parameters found in the
$C_{ijkl}$ operators, in addition to the various SUSY parameters. Recall
that the {\bf 10} mass parameter $x_0$ must be ${\cal O}(1)$ to allow
the SUSY Higgs fields to be light; the remaining mixing parameters are
functions of many undetermined GUT-scale masses and couplings found in
the full superpotential, the details of which can be seen in 
\cite{aulgarg}. We simply take the $x$s and $y$s as the free
parameters as discussed above eq.\,(\ref{eq:Cs}).

Ideally, one would find that all partial lifetimes are clear of the
experimentally determined lower limits, given in Table
\ref{table:explims}, for arbitrary values $x,y \in (0,1)$, in the
absence of unlucky enhancements. It is well-known however that some
careful cancellations are required to achieve agreement for a typical
GUT model, often of even several orders of magnitude ({\it e.g.},
$\mathrm{C}^{\cal A} = -\mathrm{C}^{\cal B} + {\cal O}(10^{-3\hbox{-}4})$).
\begin{table}[b]
\begin{center}
  \begin{tabular}{||c|c||}\hline\hline
    decay mode & $\tau$ exp lower limit (yrs) \\ \hline 
    $p \rightarrow K^+ \bar\nu$   &  $6.0 \!\times\! 10^{33}$ \\
    $p \rightarrow K^0 e^+$       &  $1.0 \!\times\! 10^{33}$ \\
    $p \rightarrow K^0 \mu^+$     &  $1.3 \!\times\! 10^{33}$ \\
    $p \rightarrow \pi^+ \bar\nu$ &  $2.7 \!\times\! 10^{32}$ \\
    $p \rightarrow \pi^0 e^+$     &  $1.3 \!\times\! 10^{34}$ \\
    $p \rightarrow \pi^0 \mu^+$   &  $1.0 \!\times\! 10^{34}$ \\
    \hline\hline
  \end{tabular}
  \caption{\footnotesize Experimentally determined lower limits
  \cite{babuexp} on the partial lifetimes of dominant proton decay
  modes considered in this work.}
\label{table:explims}
\end{center}
\end{table}

The Yukawa textures shown in eq.\,(\ref{eq:y}) are intended to
naturally suppress the values of some crucial $C_{ijkl}$ elements so
that the need for such extreme tuning is alleviated. In \cite{matt},
we demonstrated that some select but robust regions of parameter space
exist among the $x$s and $y$s that lead to adequate lifetimes in the
model for both type-I and II cases, but those results corresponded to
a fix CP phase value in each case.

In order to further test the ansatz here, we need to repeat the search
for parameter space for multiple fits containing all possible CP phase
values, namely the data presented in the previous section. We
specifically tested the model by comparing the most constrained $K^+\bar\nu$ 
mode partial lifetime values to the corresponding CP phase.

The search for such regions was carried out using adaptations of the
Python code using {\tt Minuit} described in \cite{matt}. Consideration
is Results for the type-I and type-II models are reported separately below.

%\begin{table}[t]
%\begin{center}
%  \begin{tabular}{||c|c||}\hline\hline
%    decay mode & baseline for $\tau$ (yrs) \\ \hline 
%    $p \rightarrow K^+ \bar\nu$   &  $3.59 \!\times\! 10^{34}$ \\
%    $p \rightarrow K^0 e^+$       &  $4.30 \!\times\! 10^{37}$ \\
%    $p \rightarrow K^0 \mu^+$     &  $1.13 \!\times\! 10^{37}$ \\
%    $p \rightarrow \pi^+ \bar\nu$ &  $1.51 \!\times\! 10^{36}$ \\
%    $p \rightarrow \pi^0 e^+$     &  $1.75 \!\times\! 10^{39}$ \\
%    $p \rightarrow \pi^0 \mu^+$   &  $4.39 \!\times\! 10^{38}$ \\
%    \hline\hline
%  \end{tabular}
%  \caption{\footnotesize Hypothetical baseline partial lifetimes
%  determined using type-I solution Yukawas and $x_0 = 0.95$ with all
%  other $x_i,y_i = 0$. Note in comparing with Table \ref{table:explims} 
%  that all modes satisfy the lower limits, and most do so by several
%  orders of magnitude.}
%  \vspace{-2mm}
%\label{table:baselineI}
%\end{center}
%\end{table}

\begin{figure}[b!]
  \begin{center}
    \textbf{$\boldsymbol{K^+ \bar\nu}$ Partial Lifetime vs 
    $\boldsymbol{\delta_{\rm CP}}$ for Type I Seesaw} \\[1em]
    \includegraphics[width=7.5cm]{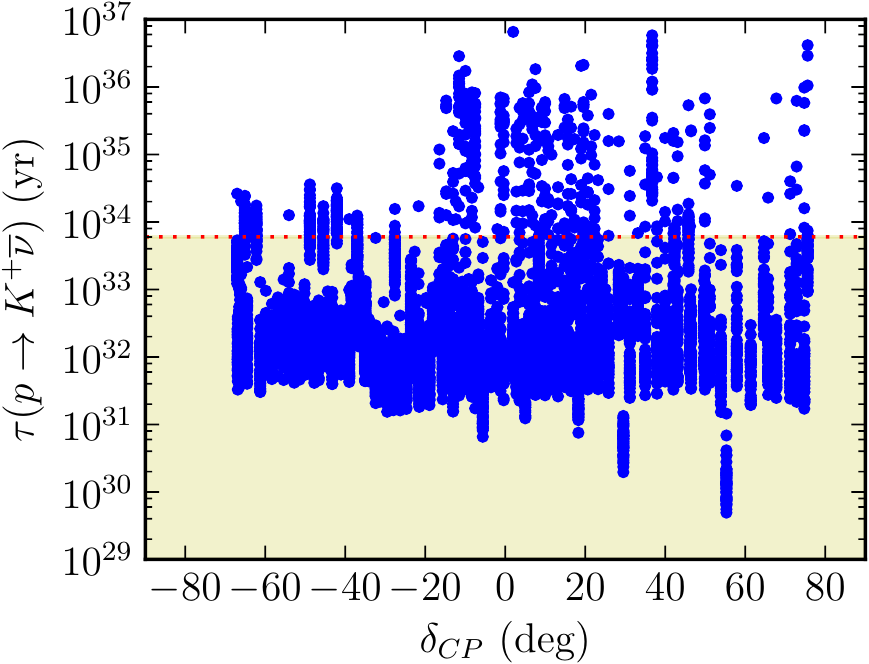}~~
    \includegraphics[width=7.5cm]{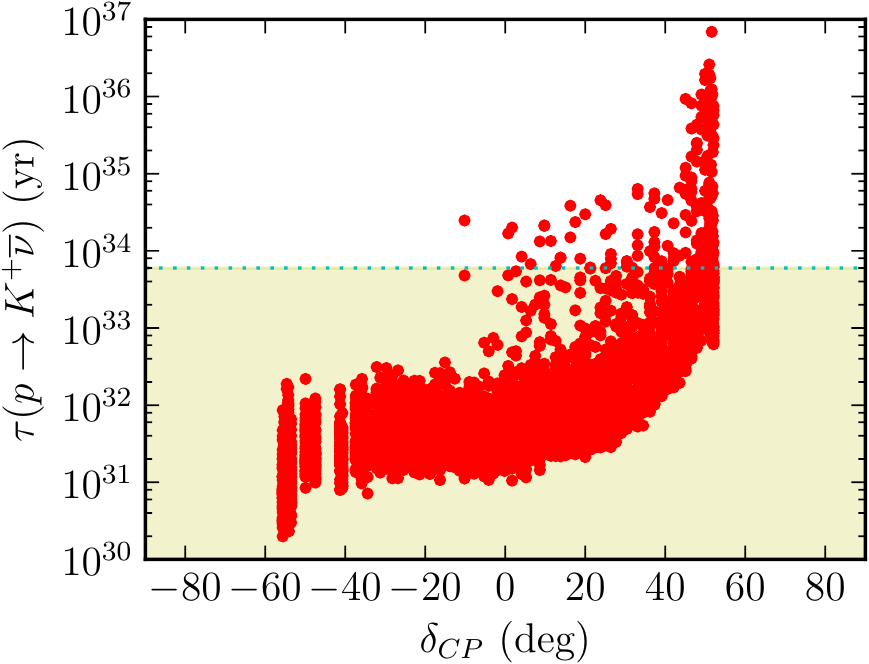}\\[1em]
    \includegraphics[width=7.5cm]{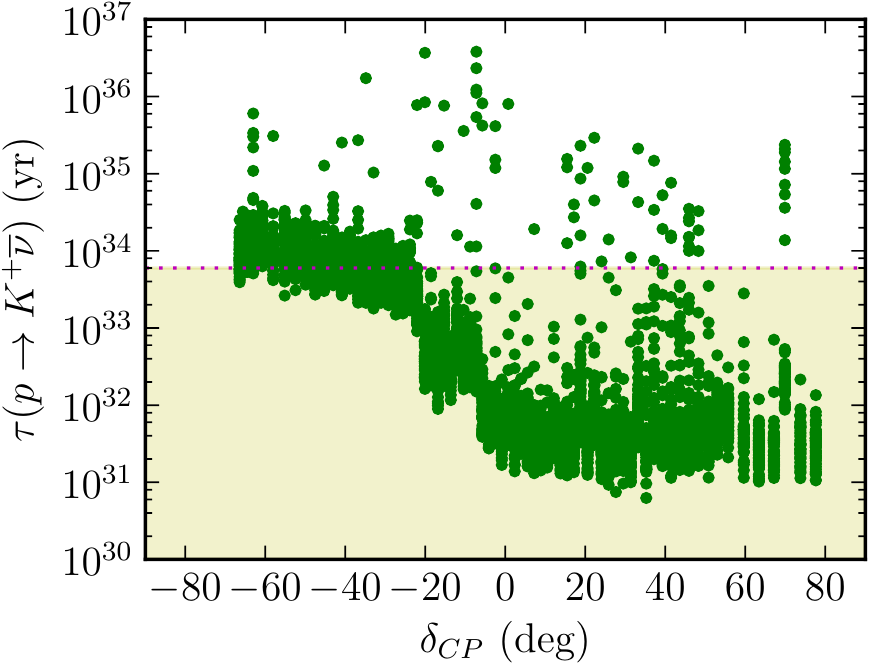}
    \caption{\footnotesize Comparisons of the Dirac CP phase angle
    to the $K^+\bar\nu$ partial lifetime, for the fermion fits
    discussed above.}
    \label{fig:KnuCPI} \vspace{-8mm}
  \end{center}
\end{figure}

\subsection{Proton Partial Lifetimes for Type I Seesaw } 
The resulting scatter plot comparing the $K^+\bar\nu$ mode partial
lifetime values to the CP phase values for the type I case is given in
Figure \ref{fig:KnuCPI}. The data coloring matches with that used 
in Section \ref{fit} for convenience. One can see that in each case, the
fraction of data for which the lifetime is sufficient is on the order
of only 20-30\%. More interesting is that each of the three initial
positions in the parameter space yields a different range of values
for $\delta_{CP}$ for which the model predicts $K\nu$ lifetimes above
the experimental threshold $6 \times 10^{33}$\,yr. Hence it seems that
a measurement of the CP phase in the future will select one region
as most favorable rather explicitly.

\subsection{Proton Partial Lifetimes for Type II Seesaw } 
\begin{figure}[b]
  \begin{center}
    \textbf{$\boldsymbol{K^+ \bar\nu}$ Partial Lifetime vs 
    $\boldsymbol{\delta_{\rm CP}}$ for Type II Seesaw} \\[1em]
    \includegraphics[width=7.5cm]{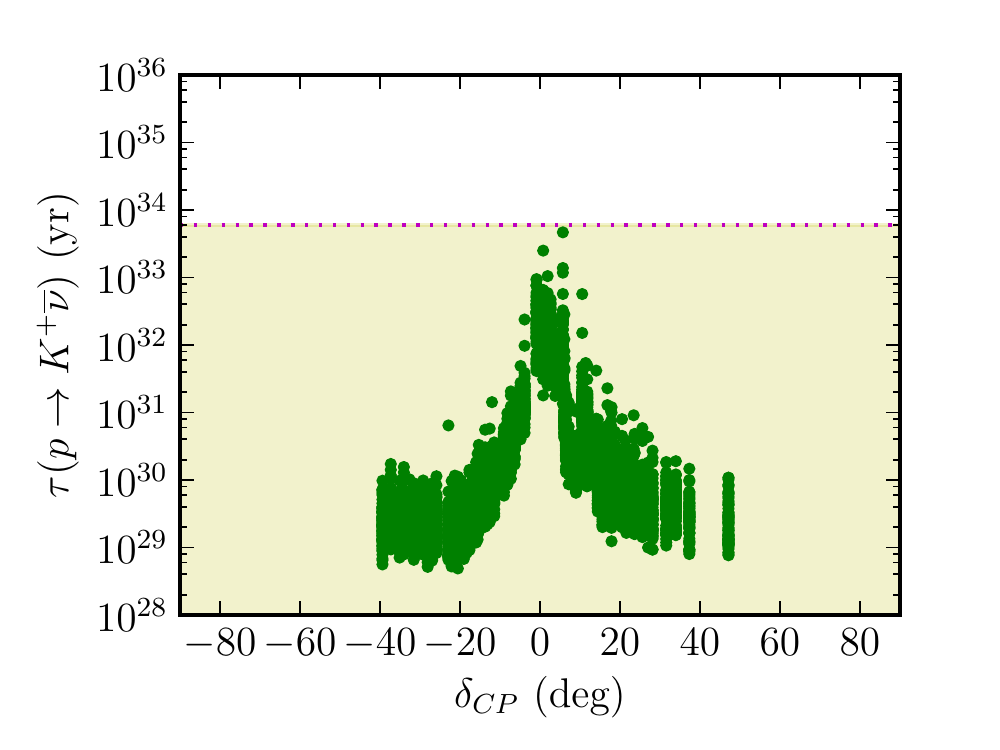}~~
    \caption{\footnotesize Comparisons of the Dirac CP phase angle
    to the $K^+\bar\nu$ partial lifetime, for the fermion fits
    discussed above.}
    \label{fig:KnuCPII} %\vspace{-10mm}
  \end{center}
\end{figure}

Despite the
heavy constraints put on the Type II case by the fermion sector
fitting in Section \ref{fit}, we report here the corresponding $K\nu$
lifetime for that data as well. The resulting scatter plot comparing
the $K^+\bar\nu$ mode partial lifetime values to the CP phase values
for the type II case is given in Figure \ref{fig:KnuCPII}. Here one
can see several unfavorable features.  First, $K\nu$ lifetime values
for all data are short of the experimental limit (for SUSY breaking
scale of 5 TeV). Recall that the analysis in \cite{matt}, which was
more selective in searching for initial positions in the $x\,
\hbox{-}\, y$ parameter space, did find a moderately-sized region of
parameter space in this model for which the lifetime was adequate, so
in that sense, this result is not a complete condemnation of the
model; however, that individual fit had $\delta_{CP} \sim -12^\circ$
and $\sum\chi^2 \sim 6$. The second, and more condemning feature of
Figure \ref{fig:KnuCPII} is that the regions of {\it lowest} lifetime
values are those which correspond to the {\it lowest} chi-square
values; hence {\it sufficient $K\nu$ lifetime and sufficient fermion
fit are in contention with one another}. Finally note that {\it the
only region with feasibly adequate $K\nu$ lifetime corresponds to
$\delta_{CP} \sim 0$, {\it i.e.}, no CP violation in the lepton
sector.} 

Hence we can further emphasize the imminent failure of the type-II-dominant
case determined by this analysis: experimental discovery of any large
CP-violating phase in the lepton sector, either positive or negative,
will rule out the type-II version of this model. Additionally, any
further raising of the experimental limit for the $K^+\bar\nu$ partial
lifetime by forthcoming analyses from DUNE, HyperK, T2K, etc would
similarly rule out this version of the model if SUSY breaking scale is 5 TeV.
\newpage

\section{Summary and Conclusion} \label{conc} To summarize, we have
considered a renormalizable SO(10) model that includes a complete and
predictive neutrino sector, with seesaw scale emerging from the
constraints of coupling unification. This particular model
additionally has all fermion masses fit by a specific texture of
Yukawa couplings that reduce the fine tuning among different
contributions to proton decay.

This model is currently of specific interest because it can be tested
by coming experiments designed to measure leptonic CP violation as
well as those to measure proton decay. It is particularly interesting
that for the current limits on proton lifetime, this model allows a
supersymmetry breaking scale in the low-TeV range, which can be tested
at the LHC and hadron colliders being planned. Thus our analysis here
explored the predictions of the model pertaining to the Dirac CP phase
and proton lifetime.

In the Type I seesaw case, we find a robust fermion sector which can
support nearly any value for the Dirac CP phase; additionally we find
a correlation between the phase value and partial lifetime for the
proton decay mode $p\to K^+\bar\nu$ that will allow the parameter
space of the model to be reduced if measurements for these observables
emerge.

In contrast, we find that possible phase values for the Type II case
are strongly limited to small values only, and that phase correlation
with $K^+\bar\nu$ partial lifetime further restricts the model into
nearly vanishing regions of parameter space. Any discovery of leptonic
CP violation or observed increase in proton partial lifetime limits
will rule out the Type-II-dominant mode of the model.

Finally, we have suggested that this general class of models with
${\bf 10}\oplus \overline{\bf 126}\oplus {\bf 120}$ (without any
specific Yukawa textures) has the potential to solve the strong CP
problem without the need for an axion. This makes this particular
class of renormalizable SUSY $SO(10)$ models of great interest.  Note
that our solution to strong CP does not seem to work in the absence of
supersymmetry. A detailed analysis of this strong CP solution is
currently in progress and will be the subject of a forthcoming
publication.

\section*{Acknowledgement} One of the authors (R.N.M) would like to
thank K. S. Babu for discussions. This work is supported by the US
National Science Foundation under Grant No. PHY1620074.


\begin{thebibliography}{99}

\bibitem{seesaw} P. Minkowski, Phys. Lett. {\bf B67}, 421 (1977); T.
  Yanagida in Workshop on Unified Theories, KEK Report 79-18, p. 95
  (1979); M. Gell-Mann, P. Ramond and R. Slansky, Supergravity, p.
  315; Amsterdam: North Holland (1979); S. L. Glashow, 1979 Cargese
  Summer Institute on Quarks and Leptons, p. 687; New York: Plenum
  (1980); R. N. Mohapatra and G. Senjanovic, Phys. Rev. Lett. {\bf
  44}, 912 (1980).
  
\bibitem{king} For recent reviews and references to literature, see
    S.~F.~King,
  %``Unified Models of Neutrinos, Flavour and CP Violation,''
  Prog.\ Part.\ Nucl.\ Phys.\  {\bf 94}, 217 (2017)
 % doi:10.1016/j.ppnp.2017.01.003
  [arXiv:1701.04413 [hep-ph]]; S.~F.~King and C.~Luhn,
  %``Neutrino Mass and Mixing with Discrete Symmetry,''
  Rept.\ Prog.\ Phys.\  {\bf 76}, 056201 (2013)
 % doi:10.1088/0034-4885/76/5/056201
  [arXiv:1301.1340 [hep-ph]].
  
\bibitem{raby} S.~Raby,
  %``Supersymmetric Grand Unified Theories : From Quarks to Strings
  %via SUSY GUTs,''
  Lect.\ Notes Phys.\  {\bf 939}, 1 (2017);  P.~Nath and R.~M.~Syed,
  %``A Stronger Case for Superunification Post Higgs Boson
  %Discovery,''
  Phys.\ Scripta {\bf 92}, no. 12, 124005 (2017).

\bibitem{so10} H. Georgi, in: Particles and Fields 1974, ed. C.E.
  Carlson (AIP, NY, 1975) p. 575.; H. Fritzsch and P. Minkowski,
  Ann. Phys. {\bf 93}, 193 (1975).

\bibitem{BM} K. S. Babu and R. N. Mohapatra, Phys. Rev. Lett. {\bf
  70}, 2845 (1993) [hep-ph/9209215].

\bibitem{other}  T. Fukuyama and N. Okada, JHEP 0211, 011 (2002)
  [hep-ph/0205066]; B. Bajc, G. Senjanovic and F. Vissani, Phys. Rev.
  Lett. {\bf 90}, 051802 (2003) [hep-ph/0210207]; Phys. Rev. {\bf D
  70}, 093002 (2004) [hep-ph/0402140]; H. S. Goh, R. N. Mohapatra and
  S. -P. Ng, Phys. Lett. {\bf B 570}, 215 (2003) [hep-ph/0303055];
  Phys. Rev. {\bf D 68}, 115008 (2003) [hep-ph/0308197]; B. Dutta, Y.
  Mimura and R. N. Mohapatra, Phys. Rev. {\bf D 69}, 115014 (2004)
  [hep-ph/0402113]; K. S. Babu and C. Macesanu, Phys. Rev. {\bf D 72},
  115003 (2005) [hep-ph/0505200]; T. Fukuyama, A. Ilakovac, T.
  Kikuchi, S. Meljanac and N. Okada, Eur. Phys. J. C {\bf 42}, 191 (2005)
  [hep-ph/0401213]; Phys. Rev. {\bf D 72}, 051701 (2005) [hep-ph/0412348];
  B. Bajc, A. Melfo, G. Senjanovic and F. Vissani, Phys. Rev. {\bf D 70},
  035007 (2004) [hep-ph/0402122]; Phys. Lett. {\bf B 634}, 272 (2006)
  [hep-ph/0511352]; C. S. Aulakh and A. Girdhar, Nucl. Phys. {\bf B 711},
  275 (2005) [hep-ph/0405074];  T. Fukuyama, A. Ilakovac, T. Kikuchi,
  S. Mel- janac and N. Okada, J. Math. Phys. {\bf 46}, 033505 (2005)
  [hep-ph/0405300]; S. Bertolini, T. Schwetz and M. Malinsky, Phys.
  Rev. {\bf D 73}, 115012 (2006) [hep-ph/0605006].

\bibitem{120} B.~Dutta, Y.~Mimura and R.~N.~Mohapatra,
  %``Neutrino masses and mixings in a predictive SO(10) model with CKM
  %CP violation,''
  Phys.\ Lett.\ B {\bf 603}, 35 (2004)
  %doi:10.1016/j.physletb.2004.09.076
  [hep-ph/0406262];  S.~Bertolini, M.~Frigerio and M.~Malinsky,
  %``Fermion masses in SUSY SO(10) with type II seesaw: A Non-minimal predictive scenario,''
  Phys.\ Rev.\ D {\bf 70}, 095002 (2004) [hep-ph/0406117].
 
  
\bibitem{120other} W.~Grimus and H.~Kuhbock,
  %``A renormalizable SO(10) GUT scenario with spontaneous CP violation,''
  Eur.\ Phys.\ J.\ C {\bf 51}, 721 (2007)
  doi:10.1140/epjc/s10052-007-0324-5
  [hep-ph/0612132]; W.~Grimus and H.~Kuhbock,
  %``Fermion masses and mixings in a renormalizable SO(10) x Z(2) GUT,''
  Phys.\ Lett.\ B {\bf 643}, 182 (2006)
  %doi:10.1016/j.physletb.2006.10.038
  [hep-ph/0607197].
 
\bibitem{RPV} R.~N.~Mohapatra,
  %``New Contributions to Neutrinoless Double beta Decay in
  %Supersymmetric Theories,''
  Phys.\ Rev.\ D {\bf 34}, 3457 (1986); A.~Font, L.~E.~Ibanez and
  F.~Quevedo,
  %``Does Proton Stability Imply the Existence of an Extra Z0?,''
  Phys.\ Lett.\ B {\bf 228}, 79 (1989); S.~P.~Martin,
  %``Some simple criteria for gauged R-parity,''
  Phys.\ Rev.\ D {\bf 46}, R2769 (1992).
  
\bibitem{aulakh} C.S. Aulakh and R.N. Mohapatra, Phys. Rev. {\bf D28},
  217 (1983); T.E. Clark, T.K.Kuo, and N.Nakagawa, Phys. lett.  {\bf
  115B}, 26(1982).

\bibitem{theta13} H. S. Goh, R. N. Mohapatra and S. -P. Ng, Phys.
  Lett. {\bf B 570}, 215 (2003) [hep-ph/0303055]; K. S. Babu and C.
  Macesanu, Phys. Rev. D 72, 115003 (2005) [hep-ph/0505200].

\bibitem{type2} J. Schechter and J. W. F. Valle, Phys. Rev. D 22, 2227
  (1980); G. Lazarides, Q. Shafi and C. Wetterich, Nucl. Phys. B 181,
  287 (1981); R. N. Mohapatra and G. Senjanovic, Phys. Rev. D 23, 165
  (1981).

\bibitem{BSV} B. Bajc, G. Senjanovic and F. Vissani, Phys. Rev. Lett.
  {\bf 90}, 051802 (2003) [hep-ph/0210207]; Phys. Rev. {\bf D 70},
  093002 (2004) [hep-ph/0402140].

\bibitem{pdk} J.~C.~Pati and A.~Salam,
  %``Lepton Number as the Fourth Color,''
  Phys.\ Rev.\ D {\bf 10}, 275 (1974)
  Erratum: [Phys.\ Rev.\ D {\bf 11}, 703 (1975)]; H.~Georgi and S.~L.~Glashow,
  %``Unity of All Elementary Particle Forces,''
  Phys.\ Rev.\ Lett.\  {\bf 32}, 438 (1974).

\bibitem{pdecay} H. S. Goh, R. N. Mohapatra, S. Nasri and S. P. Ng,
  Phys. Lett. B 587, 105 (2004) [hep-ph/0311330]; T. Fukuyama, A.
  Ilakovac, T. Kikuchi, S. Meljanac and N. Okada, JHEP 0409, 052
  (2004) [hep-ph/0406068]; M. Severson, Phys. Rev. D 92, no. 9, 095026
  (2015) [arXiv:1506.08468 [hep-ph]]; arXiv:1601.06478 [hep-ph];
T.~Fukuyama, K.~Ichikawa and Y.~Mimura,   Phys.\ Lett.\ B {\bf 764}, 114 (2017); 

\bibitem{babu} K. S. Babu, B. Bajc and S. Saad, See talk at the INT
  workshop on "Neutron-anti-neutron oscillation", October, 2017.

\bibitem{DMM} B. Dutta, Y. Mimura and R. N. Mohapatra, Phys. Rev.
  Lett. {\bf 94}, 091804 (2005) [hep-ph/0412105]; Phys. Rev. {\bf D
  72}, 075009 (2005) [hep-ph/0507319].%C.~S.~Aulakh and S.~K.~Garg,
  %``The New Minimal Supersymmetric GUT : Spectra, RG analysis and
  %Fermion Fits,''
  %Nucl.\ Phys.\ B {\bf 857}, 101 (2012)  [arXiv:0807.0917 [hep-ph]].
  
\bibitem{matt} M.~Severson,
  %``Neutrino Sector and Proton Lifetime in a Realistic SUSY SO(10)
  %Model,''
  Phys.\ Rev.\ D {\bf 92}, no. 9, 095026 (2015)
  %doi:10.1103/PhysRevD.92.095026
  [arXiv:1506.08468 [hep-ph]].
  
\bibitem{Nova} P.~Adamson {\it et al.} [NOvA Collaboration],
  %``Constraints on Oscillation Parameters from $\nu_e$ Appearance and
  %$\nu_\mu$ Disappearance in NOvA,''
  Phys.\ Rev.\ Lett.\  {\bf 118}, no. 23, 231801 (2017)
  %doi:10.1103/PhysRevLett.118.231801
  [arXiv:1703.03328 [hep-ex]]; K. Abe et al. [T2K Collaboration],
  Phys. Rev. {\bf D 91}, no. 7, 072010 (2015) [arXiv:1502.01550
  [hep-ex]].

\bibitem{DUNE} R. Acciarri et al. (DUNE) (2015), 1512.06148; 1601.02984

\bibitem{T2K} K. Abe et al. [T2K Collaboration], Phys.Rev.D 91,no.7,
  072010 (2015).
%doi:10.1103/PhysRevD.91.072010, 1502.01550.

\bibitem{hyperK} E.~Kearns {\it et al.} [Hyper-Kamiokande Working
  Group],
  %``Hyper-Kamiokande Physics Opportunities,''
  arXiv:1309.0184 [hep-ex]; M.~Yokoyama [Hyper-Kamiokande Proto
  Collaboration],
  %``The Hyper-Kamiokande Experiment,''
  arXiv:1705.00306 [hep-ex]; D.~R.~Hadley [Hyper-K Collaboration],
  %``The Hyper-K experiment: Present and R&D for the next decade,''
  Nucl.\ Instrum.\ Meth.\ A {\bf 824}, 630 (2016).
  
\bibitem{JUNO} F. An et al. (JUNO Collaboration), J.  Phys. G (Nucl.
  and Part. Phys.) {\bf 43}, id. 030401 (2016); A.~Giaz [JUNO
  Collaboration],
  %``Status and perspectives of JUNO experiment,''
  PoS EPS {\bf -HEP2017}, 108 (2018).
  
\bibitem{aulgarg} C.~S.~Aulakh and S.~K.~Garg,
  %``The New Minimal Supersymmetric GUT : Spectra, RG analysis and
  %Fermion Fits,''
  Nucl.\ Phys.\ B {\bf 857}, 101 (2012)
    [arXiv:0807.0917 [hep-ph]];  Z.~Y.~Chen, D.~X.~Zhang and X.~Z.~Bai,
  %``Couplings in Renormalizable Supersymmetric SO(10) Models,''
  Int.\ J.\ Mod.\ Phys.\ A {\bf 32}, no. 36, 1750207 (2017)
  %doi:10.1142/S0217751X17502074
  [arXiv:1707.00580 [hep-ph]].

\bibitem{strongCP} M. A. B. Beg and H.-S. Tsao,  Phys. Rev. Lett.
  {\bf 41}, 278 (1978); R. N. Mohapatra and G. Senjanovic, Phys.
  Lett. {\bf 79B}, 283 (1978); A. E. Nelson, Phys.Lett. {\bf B136},
    387 (1984); S. M. Barr,  Phys.Rev.Lett. {\bf 53}  329 (1984); K.
        S. Babu and R. N. Mohapatra, Phys. Rev. {\bf D 41}, 1286
        (1990); S.~M.~Barr, D.~Chang and G.~Senjanovic,
  %``Strong CP problem and parity,''
  Phys.\ Rev.\ Lett.\  {\bf 67}, 2765 (1991); R. Kuchimanchi, Phys.
  Rev. Lett. {\bf 76}, 3486 (1996) [hep-ph/9511376]; R. N. Mohapatra
  and A. Rasin, Phys. Rev. Lett. {\bf 76}, 3490 (1996)
  [hep-ph/9511391]; R. N. Mohapatra and A. Rasin, Phys. Rev. {\bf D
  54}, 5835 (1996) [hep-ph/9604445]; L.~J.~Hall and K.~Harigaya,
  %``Implications of Higgs Discovery for the Strong CP Problem and
  %Unification,''
  arXiv:1803.08119 [hep-ph].
    
\bibitem{matt2} P.~S.~Bhupal Dev, B.~Dutta, R.~N.~Mohapatra and M.~Severson,
  %``$\theta_{13}$ and Proton Decay in a Minimal $SO(10) \times S_4$
  %model of Flavor,''
  Phys.\ Rev.\ D {\bf 86}, 035002 (2012)
  doi:10.1103/PhysRevD.86.035002
  [arXiv:1202.4012 [hep-ph]].
  
\bibitem{nonsusy} A. S. Joshipura and K. M. Patel, Phys. Rev. D 83,
  095002 (2011) [arXiv:1102.5148 [hep- ph]]; F. Buccella, D.
  Falcone, C. S. Fong, E. Nardi and G. Ricciardi, Phys. Rev. D 86
  (2012) 035012 [arXiv:1203.0829 [hep-ph]]; Y. Mambrini, N. Nagata,
  K. A. Olive, J. Quevillon and J. Zheng, Phys. Rev. D 91, no. 9,
  095010 (2015) [arXiv:1502.06929 [hep-ph]]; K.~S.~Babu and
  S.~Khan,
  %``Minimal nonsupersymmetric $SO(10)$ model: Gauge coupling
  %unification, proton decay, and fermion masses,''
  Phys.\ Rev.\ D {\bf 92}, no. 7, 075018 (2015)
  %doi:10.1103/PhysRevD.92.075018
  [arXiv:1507.06712 [hep-ph]].
 
\bibitem{pdg} K.A. Olive {\it et al.} (Particle Data Group), Chin.
  Phys. C, {\bf 38}, 090001 (2014). 

\bibitem{gavela-king} M. B. Gavela {\it et al.} Nucl. Phys. {\bf
  B312, 2} (1989) 269.

\bibitem{fukugita} S. Aoki {\it et al.} Phys. Rev. {\bf D62}, 014506
  (2000), [arXiv:hep-lat/9911026].

\bibitem{claudson} M. Claudson, M. B. Wise, L. J. Hall. Nucl. Phys.
  {\bf B195} (1982) 297.

\bibitem{donoghue} J. F. Donoghue, E. Golowich. Phys. Rev. {\bf D26}
  (1982) 3092.

\bibitem{hisano} M. Matsumoto, J. Arafune, H. Tanaka, K. Shiraishi.
  Phys. Rev. {\bf D46} (1992) 3966; J. Hisano, H. Murayama, T.
  Yanagida. Nucl. Phys. {\bf B402} (1993) 46.

\bibitem{minuit} F. James and M. Roos, Computer Physics
  Communications. {\bf 10} (1975) 343;
  http://seal.web.cern.ch/seal/MathLibs/5 10/Minuit2/html/

\bibitem{python} G. van Rossum and F.L. Drake (eds). Python Reference
  Manual, Virginia: Python Labs (2001); http://www.python.org

\bibitem{das} C.R. Das, M.K. Parida. Eur. Phys. Journal {\bf C20} (2001)
  121 [arXiv:hep-ph/0010004].

\bibitem{bora} K. Bora. {\it Horizon, A Journal of Physics}, {\bf 2}
  (2013) ISSN 2250-0871, [arXiv:1206.5909 [hep-ph]].
  
\bibitem{poko} T. Blazek, S. Raby, S. Pokorski. Phys. Rev. {\bf D52},
  4151 (1995) [hep-ph/9504364].

\bibitem{babuexp} K. S. Babu {\it et al.} Report of the Community
  Summer Study ({\it Snowmass 2013}), Intensity Frontier -- Baryon Number
  Violation Group, [arXiv:1311.5285 [hep-ph]].

\bibitem{altarelli} G. Altarelli, G. Blankenburg, JHEP {\bf 1103}
  (2011) 133, [arXiv:1012.2697 [hep-ph]].

  \end{thebibliography}
\end{document}